\documentclass[final,5p,times,twocolumn]{elsarticle}
\journal{Physics Dark Universe}

\usepackage{ulem}
\usepackage{mathtools}

\usepackage{array}
\newcolumntype{P}[1]{>{\centering\arraybackslash}p{#1}}
\newcolumntype{M}[1]{>{\centering\arraybackslash}m{#1}}
\usepackage{graphicx}
\usepackage{lineno}
\usepackage{longtable}
\usepackage{url}
\usepackage{verbatim}
\usepackage[utf8]{inputenc}
\usepackage{makecell}
\usepackage{comment}
\usepackage[dvipsnames,table]{xcolor}
\usepackage{color}
\usepackage{cleveref}
\crefname{section}{§}{§§}
\usepackage[T1]{fontenc} 
\graphicspath{ {Figures/} }

\usepackage{hyperref}
\hypersetup{
    colorlinks=true,
    linkcolor=blue,
    filecolor=magenta,      
    urlcolor=cyan,
}

\newcommand{\jcb}[1]{\textcolor{blue}{{\small[JCB: #1]}}}
\newcommand{\masc}[1]{\textcolor{Purple}{{\small[MASC: #1]}}}
\newcommand{\mdc}[1]{\textcolor{OliveGreen}{{\small[MDORO: #1]}}}
\newcommand{\md}[1]{\textcolor{OliveGreen}{{\small #1}}}

\begin{document}

\begin{frontmatter}



\title{Sensitivity of the Cherenkov Telescope Array to dark subhalos}


\author[a,b]{Javier Coronado-Bl\'azquez}
\ead{javier.coronado@uam.es}

\author[c]{Michele Doro}
\ead{michele.doro@unipd.it}

\author[a,b]{Miguel A. S\'anchez-Conde}
\ead{miguel.sanchezconde@uam.es}

\author[a,b]{Alejandra Aguirre-Santaella}
\ead{alejandra.aguirre@uam.es}

\address[a]{Instituto de F\'isica Te\'orica UAM-CSIC,\\Universidad Aut\'onoma de Madrid, C/ Nicol\'as Cabrera, 13-15, 28049 Madrid, Spain}
\address[b]{Departamento de F\'isica Te\'orica, M-15,\\Universidad Aut\'onoma de Madrid, E-28049 Madrid, Spain}
\address[c]{Dipartimento di Fisica e Astronomia Galileo Galilei of Universit\`a degli Studi di Padova \& INFN Padova, I-35131 Padova, Italy}

\begin{abstract}
   In this work, we study the potential of the Cherenkov Telescope Array (CTA) for the detection of Galactic dark matter (DM) subhalos. We focus on low-mass subhalos that do not host any baryonic content and therefore lack any multiwavelength counterpart. If the DM is made of weakly interacting massive particles (WIMPs), these dark subhalos may thus appear in the gamma-ray sky as unidentified sources. 
   A detailed characterization of the instrumental response of CTA to dark subhalos is performed, for which we use the {\it ctools} analysis software and simulate CTA observations under different array configurations and pointing strategies, such as the scheduled extragalactic survey. This, together with information on the subhalo population as inferred from N-body cosmological simulations, allows us to predict the CTA detectability of dark subhalos, i.e., the expected number of subhalos in each of the considered observational scenarios.
   In the absence of detection, for each observation strategy we set competitive limits to the annihilation cross section as a function of the DM particle mass, that are at the level of $\langle\sigma v\rangle\sim4\times10^{-24}$ ($7\times10^{-25}$) $\mathrm{cm^3s^{-1}}$ for the $b\bar{b}$ ($\tau^+\tau^-$) annihilation channel in the best case scenario.
   Interestingly, we find the latter to be reached with no dedicated observations, as we obtain the best limits by just accumulating exposure time from all scheduled CTA programs and pointings over the first 10 years of operation. This way CTA will offer the most constraining limits from subhalo searches in the intermediate range between $\sim 1-3$ TeV, complementing previous results with \textit{Fermi}-LAT and HAWC at lower and higher energies, respectively.
\end{abstract}




\end{frontmatter}


\section{Introduction}
\label{sec:intro}
Ground-based very-high-energy (VHE) gamma-ray astronomy, using the so-called imaging air Cherenkov technique, has been profusely developed in the last decades. There are several Imaging Atmospheric Cherenkov Telescopes (IACTs) currently in operation, namely H.E.S.S. (since 2003; one 28 m and four 12 m telescopes in Namibia) \citep{Hinton2004}, MAGIC (since 2004, two 17 m telescopes in La Palma) \citep{Lorenz2004} and VERITAS (since 2007, four 12 m telescopes in Arizona) \citep{Weekes2002}.

When compared to space-based gamma-ray observatories, such as the Large Area Telescope (LAT) on board the Fermi satellite \cite{Atwood2009}, IACTs feature better energy and angular resolution, at the cost of a reduced field of view (FoV): while IACTs present a FoV of $\mathcal{O}(5^\circ)$, \textit{Fermi}-LAT observes roughly a fifth of the whole sky in a single exposure. The range of energies of both type of observatories is complementary, as satellites have significantly smaller effective areas, which degrade the sensitivity at high energies, precisely where ground-based observatories start to be competitive.

IACTs have been able to discover a significant number of VHE sources in the gamma-ray sky, producing catalogs with more than two hundred TeV sources as of today. The most populated classes are those of active galactic nuclei (AGNs), supernova remnants (SNRs) and pulsar wind nebulae (PWNe). When it comes to dark matter (DM) searches, IACT observations have been used to place competitive limits on TeV-scale DM candidates, although unable to probe the theoretically motivated annihilation thermal relic cross section value \citep{Doro2014, Gammaldi2019, fermi_magic_dsphs, Archambault2017, Abdallah2020}.

The Cherenkov Telescope Array (CTA) \cite{CTA_science_paper} is the next-generation ground-based gamma-ray observatory. Currently under construction phase, it will consist of two arrays, a Southern one located at the ESO site in the Atacama desert, Chile (24º 41' 0.34" S, 70º 18' 58.84" W), and a Northern one located at the Roque de los Muchachos in La Palma, Canary Islands, Spain (17º 53' 31.22" W, 28º 45' 43.79" N). These two arrays, in both hemispheres, will make CTA the first ground-based gamma-ray telescope able to observe almost the whole sky.
CTA will feature three types of telescopes, each with different sizes and therefore sensitive to different energy ranges: i) Large--Size Telescopes (LSTs), with 23 m diameter, sensitive to the scarce Cherenkov photons coming from gamma rays in the 20 -- 150 GeV range, ii) Medium--Size Telescopes (MSTs), 11.5 m diameter, which will observe energies between 150 GeV and 5 TeV, and iii) a large number of Small--Size Telescopes (SSTs, foreseen only in the Southern Hemisphere array) with 4 m diameter to detect the less frequent, most energetic gamma rays above 5 TeV. While the North array will be distributed within ca. 1 square kilometer in an array of 19 telescopes, CTA-South will occupy several square kilometers in an array of 99 telescopes.

As a result of this ambitious setup, all together CTA will cover energies ranging from a few GeV up to about hundreds of TeV. The large number of telescopes will also enable CTA to perform, for the first time, large surveys of extended sky regions at the highest energies\footnote{Although H.E.S.S. performed a Galactic plane survey, CTA will be able to do these kind of surveys through all the sky with extended sensitivity.}. Indeed, performing an extragalactic survey with CTA is one of the so-called \textit{key science projects} (KSPs). These represent the core topics within the CTA Consortium science program to be addressed over the first years of operations. In addition to VHE astrophysics and cosmology, these KSPs include a total of nearly $1,500$ hours devoted to the search of DM \citep{CTA_science_paper}.\footnote{We note that this observation time is not motivated by fundamental physics alone and will be used for other purposes and science cases as well; e.g., observations of the Galactic center, the Large Magellanic Cloud or the Perseus galaxy cluster.} Especially relevant in this context will be the dedicated search for annihilating DM with CTA at the Galactic Center, amply discussed in the recent Ref. \citep{Consortium2020}. It is important to stress out that the detailed specifics of CTA KSPs are still a matter of debate, and can be refined before CTA actually starts observations.

\bigskip

In addition to the already ``traditional'' astrophysical targets to search for DM annihilation (e.g., Galactic center, dwarf galaxies), there is the possibility to use CTA for less explored DM scenarios such as the one offered by the so-called Galactic \textit{dark subhalos}. These represent the less massive components of DM halo substructures in our galaxy, not massive enough to retain a baryonic, visible counterpart, as opposed to the case of the larger subhalos that host dwarf galaxy satellites. Being invisible in most of the electromagnetic spectrum, dark subhalos are hard to locate. Evidence of their passage may be obtained from their gravitational interaction on e.g., the surrounding stellar fields or by the imprint on the stellar velocity distribution \citep{Banik2018, Erkal2016,Carlberg2009,Carlberg2012,Ibata2002}. Yet, despite their unknown location, they can be excellent candidates for gamma-ray DM searches given their typical number densities, masses and distances \citep{vlii_paper,2008MNRAS.391.1685S,Brun2011,Doro2013,2017PhRvD..96f3009C,2017JCAP...04..018H} and, indeed, several groups already searched for dark subhalos in gamma-ray data and catalogs, and were able to set competitive constraints in the absence of them \citep{2010PhRvD..82f3501B,2012PhRvD..86d3504B,2012MNRAS.424L..64M,2012A&A...538A..93Z, 2012ApJ...747..121A,2012JCAP...11..050Z,2014PhRvD..89a6014B,2015JCAP...12..035B,2016JCAP...05..028S,2016ApJ...825...69M,2019JCAP...07..022A,Coronado_Blazquez2019,Coronado-Blazquez2019_2,2020Galax...8....5C}. In principle, CTA is less suitable than e.g. \textit{Fermi}-LAT to perform blind dark subhalo searches, due to its smaller FoV and more limited exposure times (orders of magnitude below that of \textit{Fermi}-LAT, which has been operating for 12 years and performs an all-sky survey every only 3 hours). A more careful discussion is deferred to Section \ref{sec:scenarios}. Yet, for this work we identify at least three different scenarios in which a subhalo may appear in the CTA FoV:

\begin{description}
    \item[EGAL scenario: the CTA planned Extragalactic Survey.] CTA plans to perform the largest sky survey ever performed by IACTs, covering a fraction of 25\% of the extragalactic ($b>5^\circ$) sky with an uniform exposure of 3~h per pointing and a total of 1000~h \citep{CTA_science_paper}. This scenario was the one also scrutinized for CTA and dark subhalos by~\citet{Huetten2016}, although several differences are noted with that work, most notably our conservative use of the Northern array, updated instrumental response functions and different N-body simulation suites.

    \item[DEEP Scenario: a deep exposure over a specific sky area.]
    CTA will be operated as an open observatory with a consistent guest observation program. Possible requests could be aimed at deep wide-field sky observation, in which a specific sky region is observed for prolonged time (say $100-150$ hours, or more).  Thanks to the possibility of divergent pointing mode \citep{Gerard2015}, CTA can scan a region of sky as narrow as about $8\times8$~deg$^2$ or as large as $20\times20$~deg$^2$ without need of repointing. 
    
    \item[EXPO scenario: overall CTA 10~years exposure.] After 10 years of continous operations, CTA is expected to have covered a considerable fraction of the sky as overall sum of its individual pointings. One could therefore use the envelope of all individual exposures and argue that potential dark subhalos could appear as serendipitous sources within the FoV of such pointings and exposures. 
\end{description}

The goal of this paper is to evaluate the sensitivity of the CTA to dark subhalos for each of these different observational strategies. We will also provide constraints on the annihilating dark matter $\langle\sigma v\rangle-m_{\chi}$ parameter space in case none of these observing scenarios would provide a detection. As shown e.g. in \citet{Coronado_Blazquez2019}, this can be done with the help of predictions from N-body cosmological simulations for a Milky-Way-size galaxy in a $\Lambda$CDM universe.

\bigskip
This paper is structured as follows. In Sec.~\ref{sec:scenarios}, we discuss the different observational search strategies. Sec.~\ref{sec:detec_and_jfact} presents the sensitivity of CTA to dark subhalos. The expected dark subhalo population and limits to the annihilation cross section are computed in Sec.~\ref{sec:constraints} and discussed in Sec.~\ref{sec:discussion}, while we conclude in Sec.~\ref{sec:conclusions}.

\section{Dark subhalo search scenarios}
\label{sec:scenarios}


Unless directly spotted by gravitational probes or detected in gamma-rays by wide FoV instruments such as those hosted in satellites (e.g. \textit{Fermi}-LAT, AGILE, HERD, DAMPE or the planned e-Astrogam or Amigo) or atmospheric shower ground-based particle detectors (e.g. HAWC, LHAASO or the planned SWGO), dark subhalos can be found only serendipitously in the CTA FoV. This can be attained in scenarios that provide either a large sky area or a large exposure: an extragalactic sky scan, a deep wide-field over a certain sky region or as unanticipated excess signals within the FoV of regular operation observations.
Note that other potentially interesting CTA programs, such as a Galactic plane survey, would be sub-optimal for subhalo searches, if one considers both the intense diffuse emission and the high density of astrophysical sources present in that region of the Galaxy.\footnote{Yet, nothing impedes that subhalo resides in the Galactic plane. In theory, by using the different spectral features of DM spectra to astrophysical ones, one could be able to disentangle them in case of detection. The prospects for such search are left for future works.} 

In the following, we discuss in more details the putative detection scenarios identified above.

\subsection{EGAL}
\begin{figure}[!ht]
\centering
\includegraphics[width=1\linewidth]{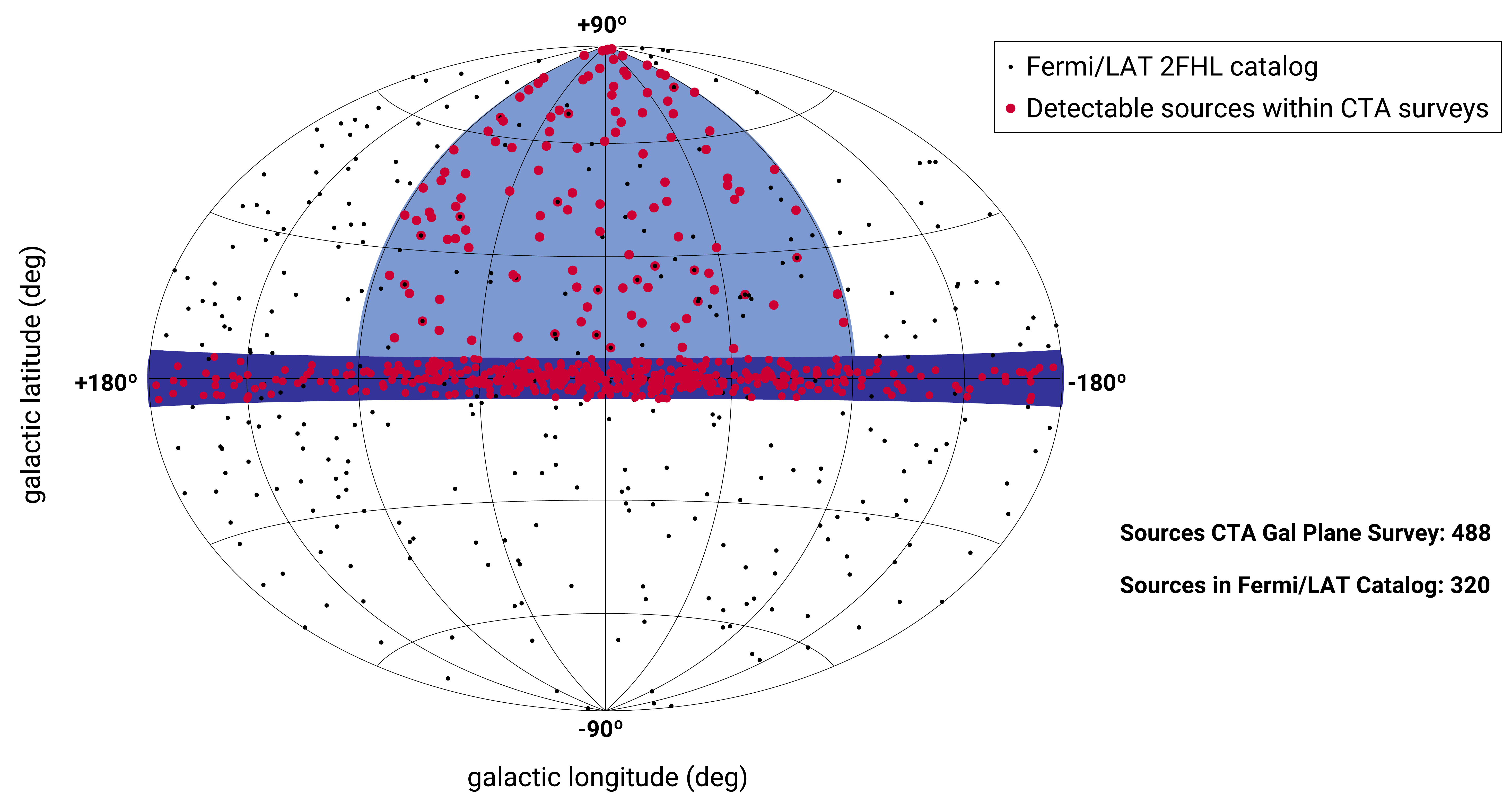}
\caption{Skymap of the EGAL Survey, including also simulated sources potentially detectable by CTA. Figure adapted from Ref.~\citep{CTA_science_paper}.}
\label{fig:egal_map}
\end{figure}

The EGAL survey has been flagged as a CTA KSP by the CTA Consortium, as it will offer an unprecedented, unbiased view of the TeV extragalactic sky~\citep{CTA_science_paper}. Never before a wide sky region was observed by IACTs with constant exposure, as IACTs mostly work on target of opportunity and specific target programs. The current plan is to scan 25\% of the sky with an uniform exposure of 3~h per pointing, for a total of 1000~h over 3 years (see the survey footprint in Figure \ref{fig:egal_map}). 

The main advantage of this search strategy is having access to a wide area of the sky surveyed with homogeneous acceptance, thus allowing for unbiased results. Furthermore, being a CTA KSP, its execution is guaranteed, although the exact sky region is not fixed yet. It is then possible that within this wide area a few optimal DM targets of interest will be present -- subhalos in the form of either visible known (or yet to be discovered) dwarf satellite galaxies (dSphs) or dark subhalos. 

A drawback of this observing strategy is the relatively short exposure per pointing, with a reported average sensitivity of the full survey of around 6 mCrab ($\sim3.04\times10
^{-13}~~\mathrm{ph~cm^{-2}~s^{-1}}$) for an energy threshold of 125 GeV.

\subsection{DEEP}
A dedicated, long exposure of $100-200$~h on a specific sky region of extended FoV between $8\times8\ \textrm{deg}^2$ and $20\times20\ \textrm{deg}^2$ appears as a possible observation mode in CTA and consequently a viable scenario for dark subhalo searches with CTA as well. Currently, a specific observation program of such kind is foreseen within the CTA core program, focused on the observation of the LMC~\citep{CTA_science_paper}. However, further deep-field observations can be proposed in the future to CTA either by Consortium members or via the CTA guest observational program,which will constantly increase the fraction of CTA observation time granted to external proposals, becoming the major observational mode in few years from the CTA start of observation. For example, a deep wide-field survey of the Virgo cluster was proposed by~\citet{Doro2013}. Such regions could be observed by CTA in two modes: with either focused, patched observation, in which a large sky region is observed in ``tiles'', or divergent pointing mode, in which the FoV is enlarged by slightly misaligning the telescopes within the array, as discussed in \ref{app:divergent_pointing}.

Dark subhalos could serendipitously fall within the FoV, whereas the probability of detection depends on the chosen sky region and its area.
As a benchmark model, we will adopt a deep wide-field of $10\times10\ \textrm{deg}^2$ with 100~h of exposure. Note though that different sky regions, different extensions, and different exposures could be required for specific science cases. 

The main drawback of this mode is that the surveyed area is relatively small, and therefore the probability to spot a dark subhalo by chance is low. Furthermore, deep, long-duration observations may be subject to higher systematics related, e.g., to the background fluctuations, inevitable over large period of times, considering the background rate depends both on the position in the sky and the atmospheric conditions during a specific data-taking (see \cite{Acharyya:2020sbj} for a detailed discussion). Such systematics effects are not taken into account in our estimates below.

\subsection{EXPO}
Regardless the specific observational proposal, during its lifetime CTA will be accumulating exposures on several directions of the sky, observation by observation. It is thus certainly possible that a dark subhalo may serendipitously appear within one of these sky regions. 

While it is not possible to predict exactly the overall CTA exposure and its sky footprint, one can make educated guesses based on available data from currently operating IACTs. In this scenario, a good modeling of the astrophysical target for which the observation was motivated is particularly critical. Only in this way it would be possible to look for any other sources hidden under potential residuals in the FoV. Another probable issue of using this third scenario for dark subhalo searches is the potentially significant loss of sensitivity a few degrees off the center of the FoV (off-axis sensitivity). For this reason, in the following we will assume the region of the FoV in which the sensitivity is expected to be almost constant.

We have based our calculation of the total, integrated CTA observation area/time as obtained after its first years of data forecasting it from the actual operations of MAGIC, an IACT currently in operation \citep{Aleksic2016, Aleksic2016a}. MAGIC is located at the Roque de los Muchachos Observatory (ORM) in the island La Palma (Spain), the same place of the northern CTA array. The MAGIC overall exposure map allows us to predict a \textit{plausible} exposure map for CTA. We defer the details of the MAGIC-to-CTA extrapolation method in \ref{app:CTA_extrapolation}.

\begin{figure}[!ht]
\centering
\includegraphics[width=1.0\linewidth]{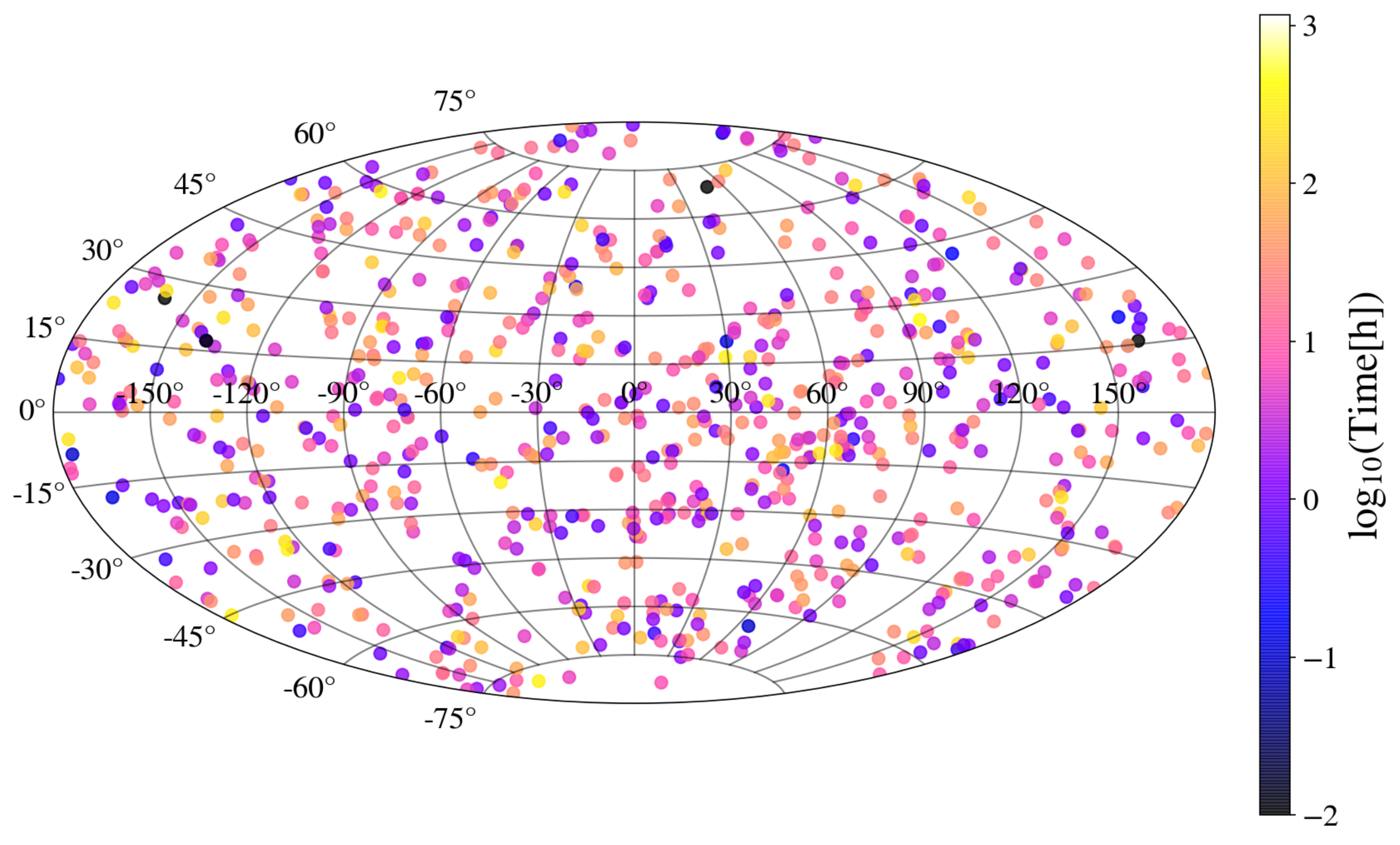}
\vfill
\includegraphics[width=0.9\linewidth]{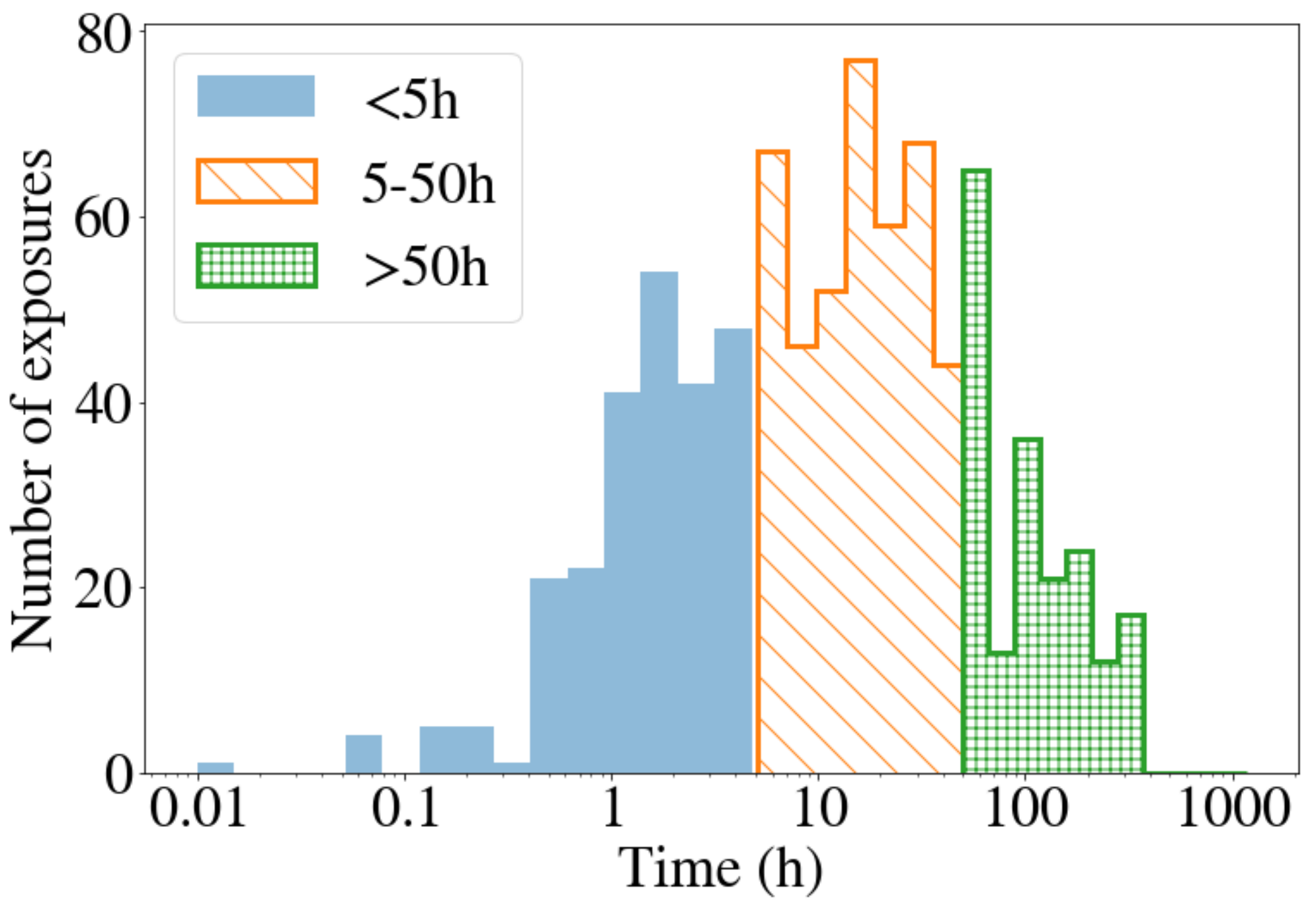}
\caption{\textbf{Upper panel}: overall 10-year exposure map for CTA in Galactic coordinates, extrapolated from 6.5 years of MAGIC stereo observations. The shown map corresponds to a random realization of our extrapolation algorithm. Color code traces the observation time per pointing. \textbf{Lower panel}: distribution of the observation time of the EXPO strategy for three bins: 0-5h, 5-50h and >50h, for the same random realization.}
\label{fig:total_exposure_map}
\end{figure}

We extrapolated MAGIC observations in stereo configuration from November 2012 to June 2019, i.e., 6.5 years of data into a projected  exposure map for CTA for 10 years of operation. Our results are shown in Figure \ref{fig:total_exposure_map}. The total CTA predicted observed area is roughly 45\% of the sky. This is almost twice that of the previously-discussed EGAL survey footprint, yet in this case unevenly sampled in terms of exposure time. 

The bottom panel of Figure \ref{fig:total_exposure_map} shows the distribution of exposure times for all the simulated CTA pointings in a random realization. One can see that $\sim$33\% of the total exposure time is devoted to short ($<5$)~h observations, $\sim$47\% of $(5-50)$~h duration dominate the observational program, and $\sim$20\% observation time is spent for very deep $(>50)$~h observations. For a fixed survey area, chances of serendipitous detection of dark subhalos are maximized during long or very long observations, as they will allow detection of DM candidates with small annihilation cross sections. Note that, in case of no-detection, DM limits can still be computed by using the full exposure map available at the time. We will do so later on in Section \ref{sec:constraints}.

\section{Dark subhalo detectability}
\label{sec:detec_and_jfact}

\subsection{CTA sensitivity to DM spectra}
\label{sec:dm_spectra}
In this section, we  evaluate the different performance of CTA to specific DM annihilation spectra. Indeed, the CTA sensitivity to DM-induced signals is expected to be different that the one for more common astrophysical ``featureless'' power-law spectra. Thus, the computation of this CTA DM sensitivity is critical should we wanted to properly address the actual chances of dark subhalo discovery.
%

For instance, for the EGAL survey, an integral sensitivity of $\sim$6 mCrab is quoted\footnote{The flux of VHE gamma rays from the Crab nebula is set to that measured by \citet{Aharonian2004}, given by $dN/dE = 2.83\times10^{-11} (E/1\ \mathrm{TeV})^{-2.62}\ \mathrm{cm}^{-2}\mathrm{s}^{-1} \mathrm{TeV^{-1}}$, while the integral flux is computed with an energy threshold of 125 GeV}. Yet, this sensitivity was computed for a Crab Nebula-like spectrum, that can be well approximated by a single power-law with a spectral index $-2.62$. Instead, DM annihilations show more complex, highly-curved spectra, strongly dependent on the mass of the WIMP and specifics of the considered annihilation channel, and with a more or less sharp cutoff at the DM mass.
To correctly compute the CTA sensitivity to DM spectra, we use the public \texttt{ctools}\footnote{\url{http://cta.irap.omp.eu/ctools/}} software \citep{ctools_paper}, v1.7.

We compute the sensitivity of CTA to dark subhalos in the EGAL survey following these steps:

\begin{enumerate}
    \item We place a simulated point-like subhalo --made of WIMPs of a given mass and annihilating via a particular channel-- in a position of the Northern Galactic hemisphere covered by the EGAL scan and close to the zenith angle of the instrumental response function (IRF)\footnote{IRFs are files containing the full characterization of a specific CTA observation mode in terms of energy resolution and bias, energy threshold, angular resolution, photon arrival direction, and effective area throughout the FoV, based on tailored optimized event cuts. More specifically, the IRFs provide the relations between the event properties as measured in the detector and the actual physical properties of the incident photon, i.e., they enclose the response of the instrument to gamma rays depending on their properties.} we use (i.e., $40^\circ$). Note that the specific coordinates of the subhalo will not change the results, as no underlying diffuse emission model is considered\footnote{This contribution is rather unknown in the TeV energy domain. As we are interested in extragalactic $\left(|b|>5^\circ\right)$ sources, a cosmic-ray induced gamma-ray diffuse emission would be negligible \citep{Gaggero2015, Gaggero2017, Cataldo2019}}. We adopt 3 hours per pointing so as to match the EGAL survey setup reported in Ref.~\citep{CTA_science_paper}. 

    \item Individual events with energies between 30 GeV up to 100 TeV are generated for a particular DM annihilation channel and WIMP mass by means of the \texttt{ctobssim} Monte Carlo generator. The input is a spectral file function obtained from PPPC4DMID \citep{Cirelli+12}, generated with \texttt{PYTHIA~8} and including electroweak corrections. Only $b\bar{b}$ and $\tau^+\tau^-$ annihilation channels are considered in this work as representative of ``soft'' and ``hard'' DM spectra, respectively, i.e., spectra exhibiting shallower and steeper cutoffs at the DM mass. For the sake of computational time, we use 29 of the 62 available masses in the PPPC4DMID tables for each channel.
    \footnote{For masses larger than $\sim10$ TeV, especially for the case of $\tau^+\tau^-$, the effect of electroweak corrections are unclear and therefore the results above this mass must be taken with caution. Concretely, these ``model-independent'' corrections assume that any effect of weakly interacting bosons radiated from the initial DM states (or from virtual internal propagators) are negligible \citep{Ciafaloni2011}. Moreover, all these radiative corrections only take into account leading order effects, while higher-order effects can potentially be included by consistently treating leading logarithms \citep{Beneke2018,Baumgart2019}. Finally, it is worth noting that an additional photon in the final state can both significantly enhance the annihilation rate \citep{Bergstroem1989} and lead to very characteristic spectral features \citep{Flores1989,Bringmann2008}, while final state gluons would only mildly modify the photon spectrum from quark final states, such as $b\bar{b}$ \citep{Bringmann2016}.}
    The events are generated using the latest IRFs, \texttt{prod3b-v2}, with the \texttt{North\_z40\_5h}\footnote{Publicly available  at \url{https://www.cta-observatory.org/science/cta-performance/}} (i.e., CTA Northern array, for a source located at a zenith angle of $40^\circ$, azimuth-averaged for 5~h observation time). 
    
    \item Once the events are simulated, we use \texttt{ctlike} to compute the detection significance via the likelihood-ratio test statistic (TS), defined as:

\begin{equation}
\label{eq:det_TS}
\mathrm{TS}=-2\,\textrm{log}\left[\frac{\mathcal{L}(H_1)}{\mathcal{L}(H_0)}\right]
\end{equation}

\noindent where $\mathcal{L}(H_0)$ and $\mathcal{L}(H_1)$ are, respectively, the likelihood functions under the null (no source) and alternative (existing source) hypotheses. The detection threshold is set to $\mathrm{TS}=25$, corresponding to about 5 standard deviations \citep{Cowan2011, Conrad2015}. Additionally, as the reconstructed event energy may differ from the true photon energy, particularly at low energies, we take into account the energy dispersion in the computation of the likelihood.\footnote{In any case, we checked that enabling and disabling this energy dispersion affects only marginally our results.}

    \item For each mass and annihilation channel, the normalization of the source flux is varied by running \texttt{ctlike} iteratively until the detection threshold is reached. We adopt a tolerance of $TS=25\pm1$ to ensure numerical convergence. The flux obtained this way is the minimum detection flux, $F_{min}$.

    \item Finally, the random seed initially used by \texttt{ctobssim} to generate the events is changed in order to randomize the distribution and create another simulated observation. We notice a smooth convergence after 10-20 realizations; nevertheless we perform 100 simulations so that we can obtain reliable uncertainty bands at 68\% and 95\% confidence levels. 
\end{enumerate}

\begin{figure}[!ht]
\centering
\includegraphics[width=1\linewidth]{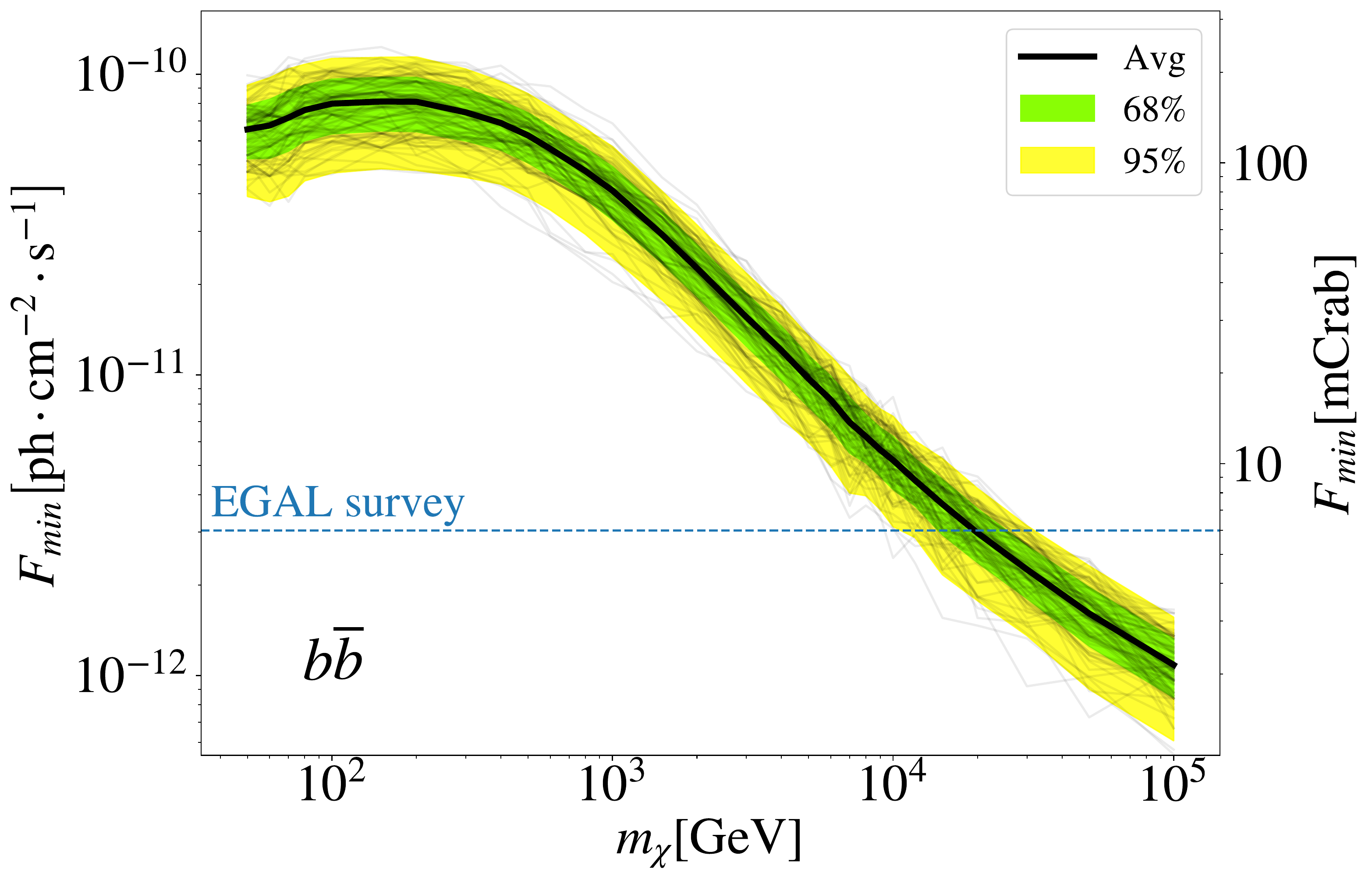}
\includegraphics[width=1\linewidth]{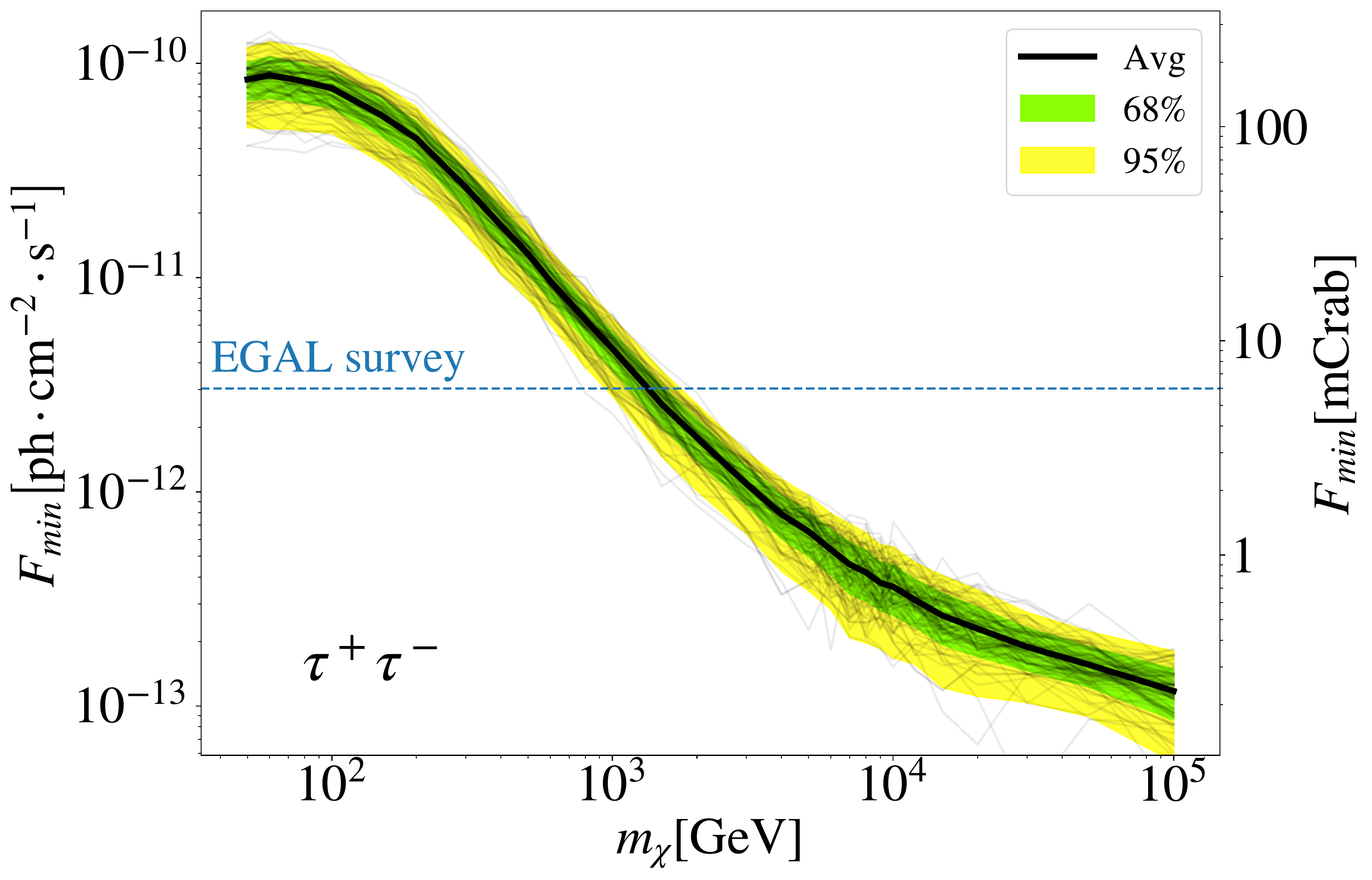}
\caption{Minimum flux $\left(F_{min}\right)$ to detect a dark subhalo annihilating to $b\bar{b}$ (top panel) and $\tau^+\tau^-$ (bottom panel) in the EGAL survey, assuming the Northern array and 3h of normal pointing observation with an energy threshold of 30 GeV. Gray lines are the individual 100 realizations, black solid line is the average value, and green and yellow bands are, respectively, the 68\% and 95\% confidence level. Right $y$ axis is expressed in units of mCrab flux. Blue dashed, horizontal line is the nominal 6 mCrab flux sensitivity of the EGAL scan.}
\label{fig:fmin}
\end{figure}

The results of our DM sensitivity computation are shown in Figure \ref{fig:fmin} for the $b\bar{b}$ and $\tau^+\tau^-$ annihilation channels. Both channels present similar behaviours. Standard Crab-based integral sensitivity overestimates the actual DM performance by a factor of $10-100$ for low DM mass (depending on the specific channel), while  the sensitivity to DM spectra reaches the standard one toward larger masses, to become even better (by a factor of up to 10 at the heaviest considered WIMP) in the case of the $\tau^+\tau^-$ channel. This is expected, as $\tau^+\tau^-$ is a harder channel compared to $b\bar{b}$. Indeed, the former peaks at $\mathrm{E_{peak}^{\tau^+\tau^-}}\sim$ $m_{\chi}/3$, while the latter peaks at $\mathrm{E_{peak}^{b\bar{b}}}\sim m_{\chi}/30$. As the best sensitivity of CTA is reached at $\sim$ $3-5$ TeV, this means that the peak of the DM annihilation spectrum coincides with CTA's best sensitivity for $m_\chi\sim 90-150$ TeV for $b\bar{b}$, and $m_\chi\sim 9-15$ TeV for $\tau^+\tau^-$. This roughly coincides with our results, as the $b\bar{b}$ sensitivity curve shown in the right panel of Figure~\ref{fig:fmin} reaches its best sensitivity at 100 TeV (the largest considered mass), and $\tau^+\tau^-$'s best sensitivity is reached and starts to flatten at masses $m_{\chi}\gtrsim 10$ TeV in the bottom panel of the same figure.

\bigskip
Note that, as pointed out above, Figure~\ref{fig:fmin} refers to the CTA sensitivity to dark subhalos under the EGAL survey setup. To compute the $F_{min}$ in either the DEEP or the EXPO scenarios, we will assume that the sensitivity scales as the square root of the observation time (using the same $40^\circ$ zenith IRFs). This saves a significant amount of computation time, as its direct calculation is computationally too expensive due to the large involved exposure times ($\sim$40 and 100 h, instead of 3h in the EGAL survey). We checked that this is a good approximation, by running some full simulations for both the 42h~ and 100~h setups. In these checks, we adopted 50h-IRFs (the available IRFs are for $0.5$, $5$ and $50$~h) and three different DM masses ($0.1$, $1$ and $10$~TeV). The square-root-of-time scaling turns out to be, indeed, an excellent representation of the actual scaling, within errors, for the three considered masses. 
In an actual observation, we note that we expect this square-root scaling to be accurate as long as there is not significant diffuse emission, a condition that is fulfilled at latitudes far from the Galactic plane.

\bigskip
Table \ref{tab:fmin} displays the results for the minimum flux $(F_\textrm{min})$ to detect a dark subhalo computed for the three different observational strategies under scrutiny, and for three benchmark WIMP masses ($0.1$, $1$ and $10$~TeV) and for both $b\bar{b}$ and $\tau^+\tau^-$ annihilation channels.

\begin{table}[h!]
\centering
\begin{tabular}{ |c|c|c|c|  }
\hline
\rowcolor[gray]{.8} 
\multicolumn{4}{|c|}{Minimum flux for detection $F_\textrm{min}$ [ph cm$^{-2}$ s$^{-1}$]} \\
\multicolumn{1}{|r|}{DM mass} & 0.1 TeV & 1  TeV & 10  TeV \\
\hline
\rowcolor[gray]{.9} 
\multicolumn{4}{|l|}{Scenario 1: EGAL (T=3h)} \\
$b\bar{b}$ & $7.98\times10^{-11}$ & $4.10\times10^{-11}$ & $5.21\times10^{-12}$\\
$\tau^+\tau^-$ & $7.65\times10^{-11}$ & $4.66\times10^{-12}$ & $3.61\times10^{-13}$\\
\hline
\rowcolor[gray]{.9} 
\multicolumn{4}{|l|}{Scenario 2: DEEP (T=100h)} \\
$b\bar{b}$ & $1.39\times10^{-11}$ & $7.14\times10^{-12}$ & $9.03\times10^{-13}$\\
$\tau^+\tau^-$ & $1.31\times10^{-11}$ & $8.02\times10^{-13}$ & $6.21\times10^{-14}$\\
\hline
\rowcolor[gray]{.9}
\multicolumn{4}{|l|}{Scenario 3: EXPO (T=42h)} \\
$b\bar{b}$ & $2.14\times10^{-11}$ & $1.07\times10^{-11}$ & $1.32\times10^{-12}$\\
$\tau^+\tau^-$ & $2.05\times10^{-11}$ & $1.22\times10^{-12}$ & $9.11\times10^{-14}$\\
\hline
\end{tabular}
\caption{Average $F_{min}$ values with an energy threshold of 30 GeV, in $\mathrm{ph\times cm^{-2}\cdot s^{-1}}$, for the three  observational strategies under discussion, and for three benchmark WIMP masses annihilating via either $b\bar{b}$ or $\tau^+\tau^-$ annihilation channels.}
\label{tab:fmin}
\end{table}

There are some caveats to these computations. First, we assume the DM subhalos to be point-like sources, while N-body cosmological simulations \citep{Coronado-Blazquez2019_2} found possible large angular sizes, up to $\mathcal{O}(5^\circ)$ for some dark subhalos according to their distance, when adopting the scale radius as the extension of the subhalo.\footnote{The choice for the scale radius is motivated because, for a Navarro-Frenk-White (NFW) \citep{NFW_paper} DM density profile, $\sim$90\% of the total annihilation flux comes from the region subtended by it.} In principle, a dedicated \textit{target-by-target} discussion would be needed, in which the specific extension is treated individually. However, applying such an approach in this study would have introduced an additional level of degeneracy with target distance, target intensity and telescope acceptance, which we prefer to avoid. In fact, understanding that the expected signal from a dark subhalo is always peaked toward its center, it is plausible to assume that a point-like search in the FoV of the three scenarios above will not provide significantly different outcomes than a specific search for moderately extended sources. This is especially true for CTA, in which the acceptance is flat throughout the inner degrees of the camera.\footnote{\url{https://www.cta-observatory.org/science/cta-performance/\#1472563544190-020879e1-468f}} 

Additionally, the $F_{min}$ computation did not take into account any underlying Galactic diffuse emission. This contribution is unknown in the TeV energy domain. Yet, as we are restricting our search to extragalactic $\left(|b|>5^\circ\right)$ sources, a cosmic-ray induced gamma-ray diffuse emission would be negligible \citep{Gaggero2015, Gaggero2017, Cataldo2019}, and, as a consequence, it would only slightly worsen the obtained $F_{min}$. A related issue can be the diffuse emission that may be generated by DM annihilation itself in the halo of the Galaxy, which may indeed outshine some of these subhalos \citep{Facchinetti2020}.

Finally, we assumed that the EGAL survey will be performed by CTA-North. However, the actual coordinates and area are still under discussion. Should the EGAL survey be performed, at least partially, with CTA-South in the end, the reached $F_{min}$ could be better than the one in our study above.\footnote{For energies higher than $\sim$1 TeV, CTA-South is roughly a factor 2 more sensitive than CTA-North, while the improvement below this energy is minimal. Therefore, one can put an upper bound of this factor 2 in the $F_{min}$ improvement, which, if relevant, will be for very heavy WIMPs, which peak at energies higher than this $\sim$1 TeV.}


\subsection{Subhalo annihilation flux}
\label{sec:j_factors}
We now continue in elucidating the number and the flux of expected dark subhalos for each specific sky region of the three scenarios described in Section~\ref{sec:scenarios}.

To perform this task, we choose the procedure to populate N-body simulations with low-mass dark clumps as described by \citet{Coronado_Blazquez2019, Coronado-Blazquez2019_2}, which are based on the Via Lactea II (VL-II) N-body cosmological simulation of a Milky Way-sized halo \citep{vlii_paper}. In the mentioned works, the original VL-II simulation was ``repopulated'' below its mass limit, due to numerical resolution, to include subhalos as light as $10^3\ \mathrm{M_{\odot}}$ (well below the original minimum subhalo mass with completeness in the simulation, of about $10^6\ \mathrm{M_{\odot}}$).\footnote{Since the subhalo mass function is not fully reproduced below one million solar masses in VL-II, and we expect those subhalos to exist in $\Lambda$CDM, we generate more of them adopting the subhalo radial distribution and an extrapolation of the subhalo mass function as found above the mass resolution of the original simulation.}
The repopulated VL-II provides very light yet very near subhalos, indeed some of them exhibiting annihilation fluxes comparable to those of more massive yet farther ones. With the repopulation algorithm of \citet{Coronado_Blazquez2019, Coronado-Blazquez2019_2}, we generate 1000 realizations of VL-II so that the distribution of subhalo fluxes can be properly drawn. From the astrophysical point of view, the DM-induced 
``brightness'' is codified into the so-called J-factor, which accounts for all the astrophysical considerations:
\begin{equation}
\label{eq:j_factor}
J=\int_{\Delta\Omega}d\psi\int_{l.o.s} \rho_{DM}^2(r(l))\,dl,
\end{equation}

\noindent where the first integral is performed along the sky line $\psi$ over the solid angle defined from the signal region ($\Delta\Omega$), the second one along the line of sight (l.o.s, $l$), and $\rho_{DM}$ is the DM density profile of the object under consideration.

In the following, we adopt for the J-factor the so-called $J_{max}$ value. This is the J-factor of the brightest subhalo, but averaged across our 1000 realizations of VL-II, and in such a way that we take the value above which 95\% of the brightest subhalo J-factor distribution is contained (i.e., out of 1000 realizations, 95\% of the time their maximum J-factors will be larger than the value we adopt). $J_{max}$ will also depend on the sky area considered, as explained below. The J-factor distribution of all realizations is drawn imposing $\mathrm{M}\leq10^8\mathrm{M_{\odot}}$ as to ensure we deal with dark subhalos (more massive subhalos will surely host a visible dSph and therefore become visible at other wavelengths \citep{Okamoto2010, Sawala2015, Sawala2017, Kelley2019}). In order to evaluate the detectability of dark subhalos, we follow the method by \citet{Coronado_Blazquez2019}.

\begin{figure}[!ht]
\centering
\includegraphics[width=1\linewidth]{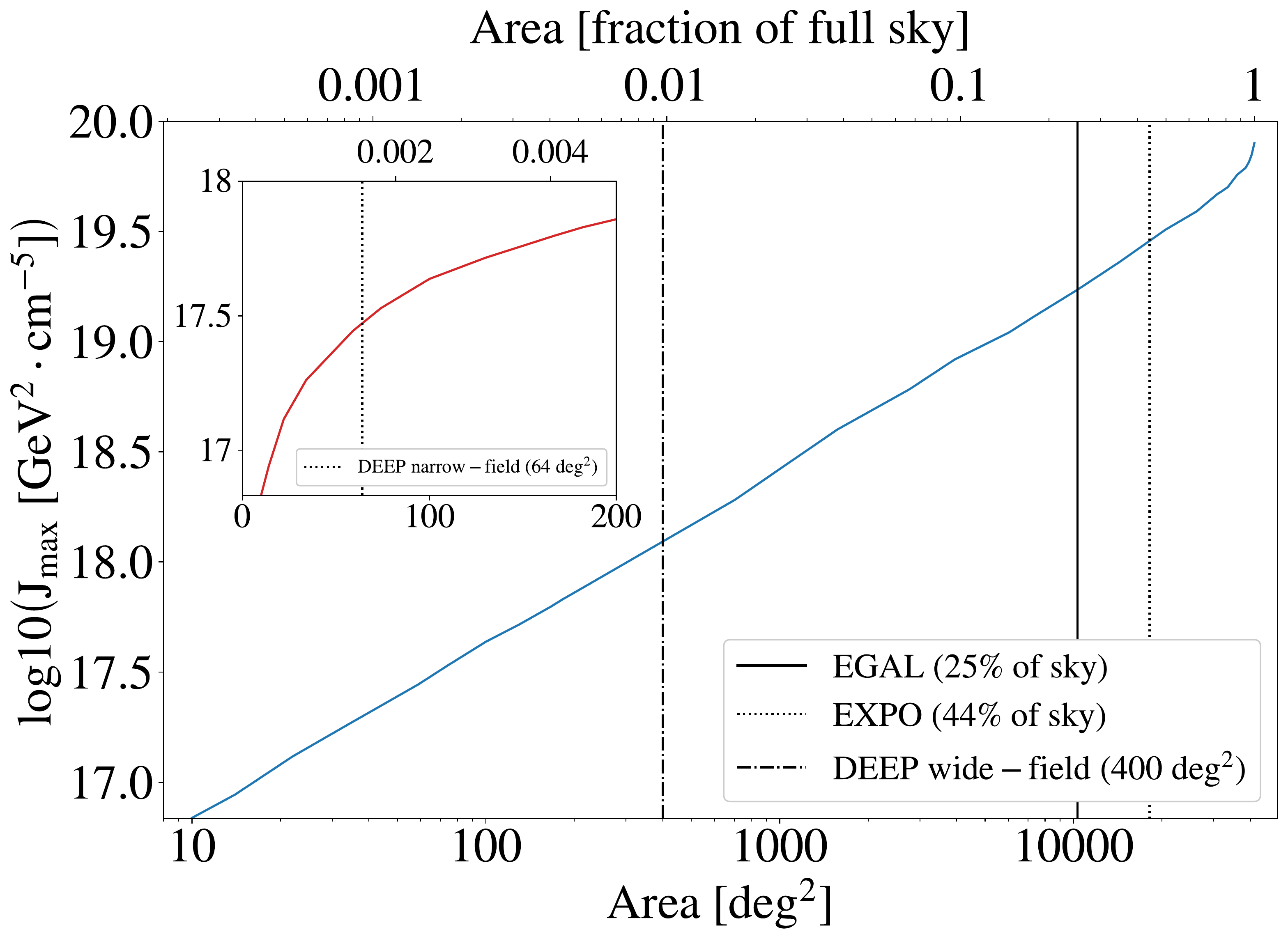}
\caption{Scaling of $J_{max}$ with the total observation area; see text for details. Only subhalos with $\mathrm{M}\leq10^8\mathrm{M_{\odot}}$ are considered, as more massive subhalos are expected to host dSphs and thus be visible. The horizontal axes are expressed in square degrees (bottom axis) and as fraction of the full sky (top axis). We show four vertical lines, respectively, at 64 square degrees ($8\times8\ \textrm{deg}^2$ DEEP scenario with narrow-field), 400 square degrees ($20\times20\ \textrm{deg}^2$ DEEP with wide-field), 25\% of the sky (EGAL) and 44\% of the sky (EXPO). The zoom panel shows the scaling for the smallest areas, in linear scale.}
\label{fig:jfact_area_scaling}
\end{figure}

As we increase the observation area, the chance of having subhalos of larger $J_{max}$ also increases, as shown in Figure \ref{fig:jfact_area_scaling}. For the case of observing 100 square degrees, $J_{max}\sim 10^{17.5}~\mathrm{GeV^2~cm^{-5}}$. When adopting e.g. 400 square degrees, subhalos with $\sim 10^{18}~\mathrm{GeV^2~cm^{-5}}$ can be expected to be observed. In the EGAL survey, with about 25\% of the sky, we can expect to have dark subhalos as bright as $J_{max}\sim 10^{19}~\mathrm{GeV^2~cm^{-5}}$. Note that, up to very large portions of the sky ($\gtrsim$80\%), the dependence of $\mathrm{log_{10}}(J_{max})$ with the area is a power law: doubling the observation area increases the logarithm of $J_{max}$ by a factor 1.6 according to the slope of the curve in Figure \ref{fig:jfact_area_scaling}. These results are in agreement with those of \cite{Dubus2013,Huetten2016}.

It is evident that, for a given total observation time, increasing the area will result in less time per pointing and a poorer sensitivity. Thus, the optimal strategy must rely on a a compromise between area and time. If we assume a square-root scaling of the sensitivity with exposure time (see Section \ref{sec:dm_spectra}), then doubling the latter will improve the sensitivity by a factor $\sqrt{2}=1.4$. This is to be compared to the factor $1.6$ mentioned before for the relation between area and $J_{max}$. Therefore, in principle it is better to increase the area at the expense of less exposure time. This will make the EXPO scenario the optimal one, as it is the one in which the observed area is the largest.

\section{Results}
\label{sec:constraints}
\subsection{Expected number of dark subhalos}
For each of the three different observation strategies discussed in Section \ref{sec:scenarios}, and making use of the subhalo annihilation flux computed in Section \ref{sec:j_factors}, it is now possible to estimate the number of expected subhalos in the CTA FoV above a certain J-factor, $J_{min}$. The results are summarized in Table \ref{tab:jmin}, where the reported uncertainties are due to the variance averaging across realizations of the VL-II simulation and across different sky positions of the observed patch.\footnote{The latter is only relevant in the case of the DEEP exposure, where the variance across sky positions is of the order of the variance across realizations. In the EGAL and EXPO strategies this uncertainty is negligible.}

\begin{table}[h!]
\centering
\begin{tabular}{c|c|c|c}
\hline
& \multicolumn{3}{c}{$N\left(\geq J_{min}\right)$} \\
\hline
$\mathrm{log_{10}}(J_{min})$ & EGAL & DEEP & EXPO\\
\hline
\rowcolor[gray]{.9} 17 & $392 \pm 18$  & $26 \pm 8$    & $728\pm 23$\\
18 & $115 \pm 11$  & $5.1 \pm 2.5$ & $206\pm 14$\\
\rowcolor[gray]{.9} 19 & $6.5 \pm 2.6$ & $0.3 \pm 0.5$ & $13 \pm 4$\\
20 & $0.3 \pm 0.5$ & $0$          & $0.5\pm 0.7$\\
\hline
\end{tabular}
\caption{Average number of subhalos above a given $J_{min}$ in the CTA FoV for the three observation strategies of Section \ref{sec:scenarios}, across 1000 realizations of the repopulated VL-II N-body simulation as described in Section \ref{sec:j_factors}. }
\label{tab:jmin}
\end{table}

Table \ref{tab:jmin} shows that there are two scenarios: EXPO and possibly EGAL, in which it will be statistically plausible to have a dark subhalo with $J=10^{20}~\mathrm{GeV^2 cm^{-5}}$ in the data. Such high value, predicted by our simulations, would compete with or even surpass the $J-$factors of the most DM dominated ultra-faint dSphs, such as Ret~II with an estimated $J-$factor of $J\sim7\times10^{19}~\mathrm{GeV^2 cm^{-5}}$~\citep{Bonnivard2015}, or that of the Large Magellanic Cloud $(\geq10^{20}~\mathrm{GeV^2cm^{-5}}$) \citep{Buckley2015}. This result is in agreement with \citet{Huetten2016} for their most optimistic scenario.\footnote{Indeed, in Ref.~\citep{Huetten2016} this would be the EGAL survey and their aggressive, so-called ``HIGH'', configuration.}

Assuming as working hypothesis a DM mass of 1 TeV, a cross-section as the thermal one $\langle \sigma v \rangle\sim 10^{-26}~ \mathrm{cm^3s^{-1}}$, the annihilation into purely $b\bar{b}$ and a target located in Northern hemisphere, a $J=10^{20}~\mathrm{GeV^2cm^{-5}}$ subhalo would give a flux at the CTA telescopes of the order of $\Phi \sim 5\times10^{-13}$~ph cm$^2$ s$^{-1}$. This can be compared with the corresponding value computed in Table~\ref{tab:fmin} that is $F_{min}=4.1\times10^{-11}$~ph cm$^2$ s$^{-1}$, roughly two orders of magnitude larger. In order to obtain a detection, this putative subhalo should require thousands of hours of exposure, far more than what achievable in the EGAL and EXPO scenarios. Similar exposures are found for all the masses and annihilation channels of Figure~\ref{fig:fmin}.

\subsection{Exclusion curve for the no-detection case}
However, even if no dark subhalo is detected by CTA, it will be still possible to set constraints to the $\langle\sigma v\rangle-m_{\chi}$ DM parameter space. The methodology is the same as in Refs.~\citep{Coronado_Blazquez2019, Coronado-Blazquez2019_2}, and is based on a comparison between the N-body simulation prediction $J_{max}$ and the gamma-ray data $F_{min}$ via:
\begin{equation}
\langle\sigma v\rangle=\frac{8\pi\cdot m_{\chi}^2\cdot F_{min}}{J_{max}\cdot N_{\gamma}}
\label{eq:master_formula}
\end{equation}

\noindent where $m_\chi$ is the DM particle mass, $F_{min}$ the instrumental sensitivity to DM as defined in Section \ref{sec:dm_spectra}, $J_{max}$ the 95\% c.l. nominal value of the brightest subhalo J-factor among the different VL-II realizations, as described in Section \ref{sec:j_factors}, and $N_{\gamma}$ is the DM spectrum for a particular annihilation channel integrated within the energy range under consideration.

\begin{figure}[!ht]
\centering
\includegraphics[width=0.9\linewidth]{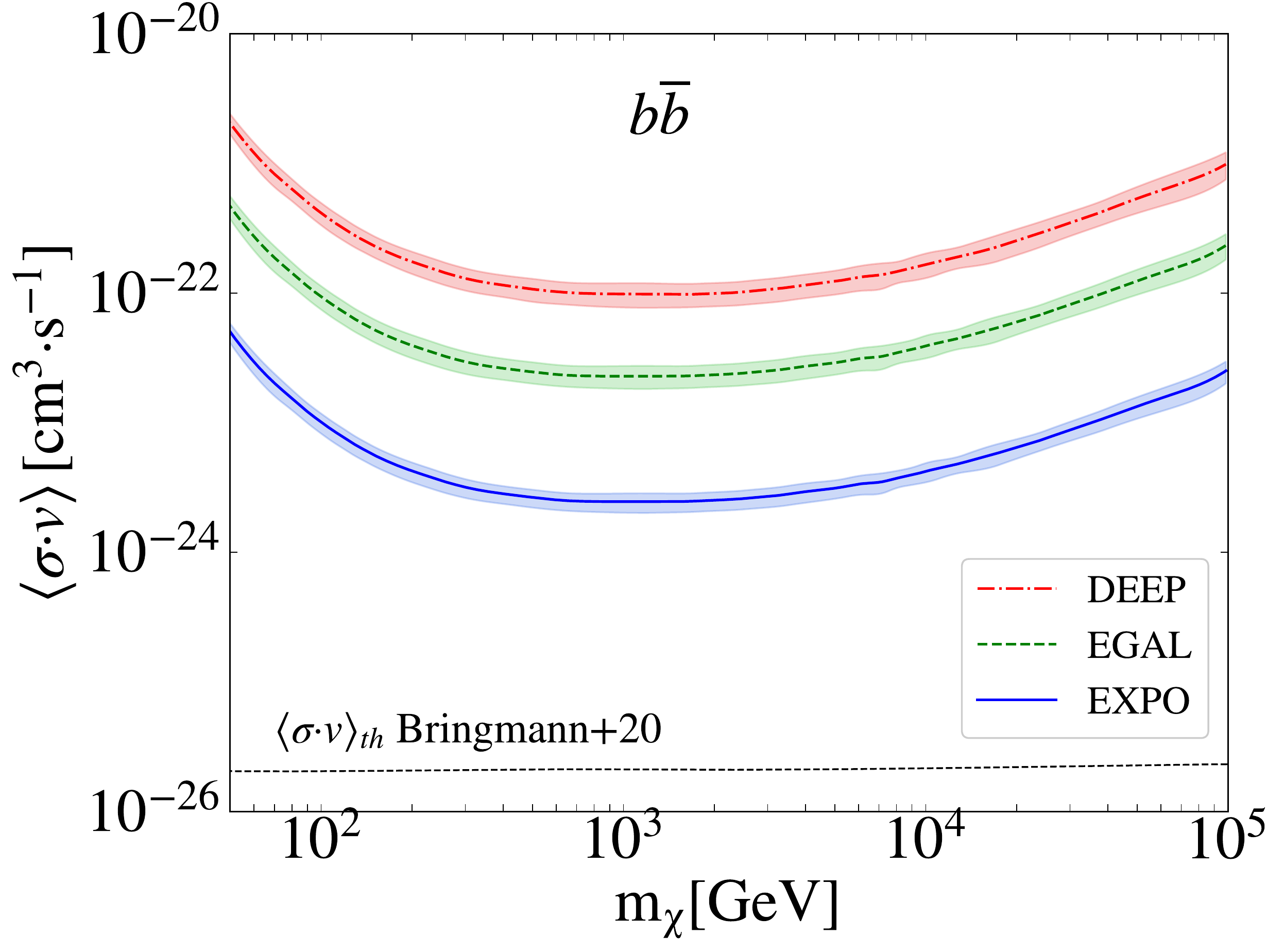}
\vfill
\includegraphics[width=0.9\linewidth]{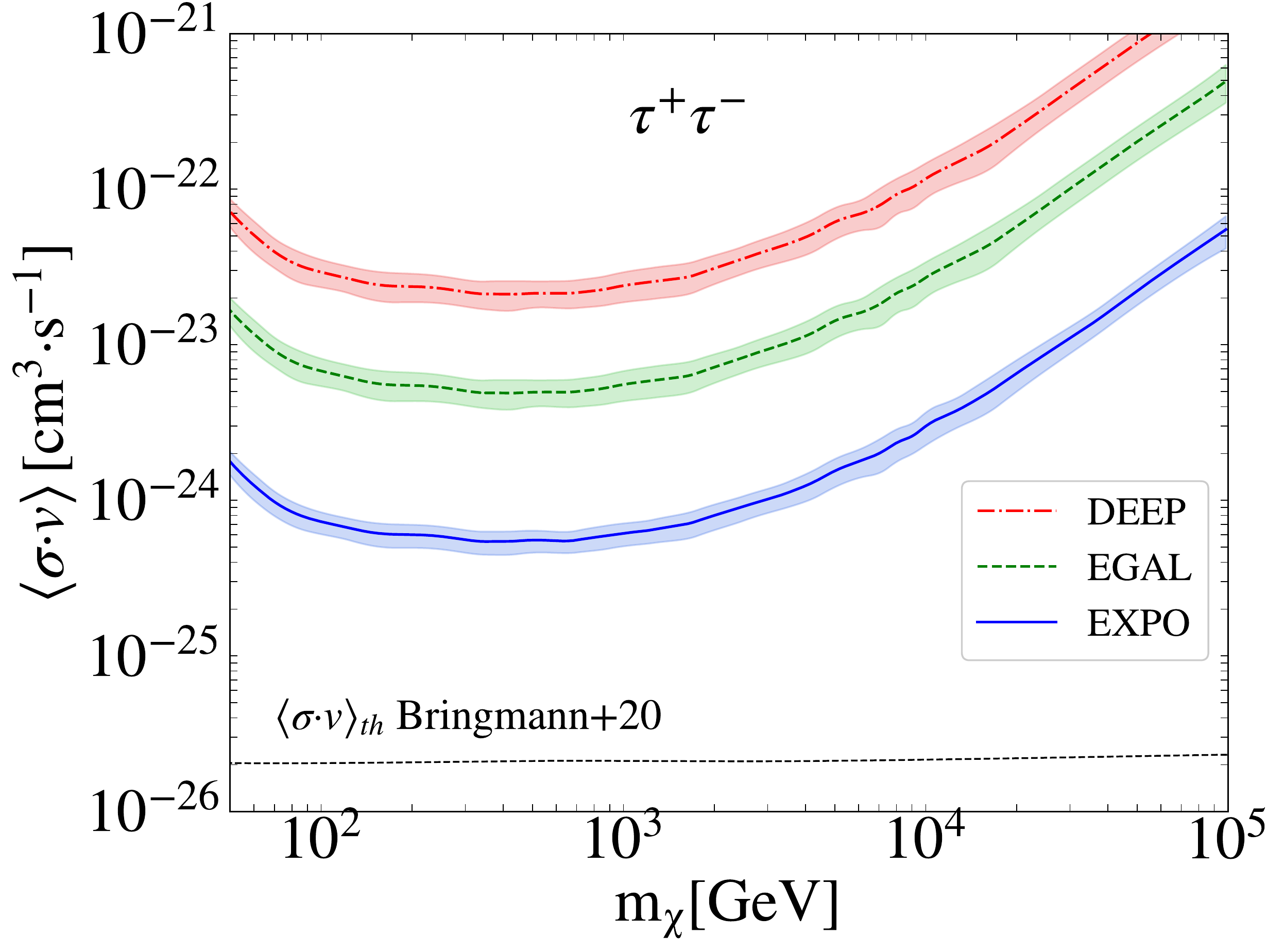}
\caption{95\% C.L. upper limits to the DM annihilation cross section for $b\bar{b}$ (top) and $\tau^+\tau^-$ (bottom) assuming no unIDs are detected by CTA under any of the three observational strategies proposed in Sec.~\ref{sec:scenarios}: a dedicated $10\times10 \ \textrm{deg}^2$, 100h deep-field, the EGAL survey, and the EXPO scenario. The dashed line represents the thermal value of the annihilation cross section \citep{Bringmann2020}. See main text for details on the uncertainty bands, which, in short, come from the uncertainty in $F_{min}$ and, in the EXPO case, the uncertainty in the total sky area extrapolation.}
\label{fig:limits_bb_tautau}
\end{figure}

Figure \ref{fig:limits_bb_tautau} shows the 95\% c.l. upper limits to the DM annihlation cross section for the $b\bar{b}$ (top panel) and $\tau^+\tau^-$ (bottom panel) annihilation channels for the three observational strategies under consideration in this work. The most stringent limits are obtained for the EXPO method, while the weakest ones are those for which the DEEP scenario is adopted.

These constraints reach their best sensitivity for masses of $\sim1$ TeV (500 GeV) for the $b\bar{b}$ ($\tau^+\tau^-$) annihilation channel, of the order of $3\times10^{-24}$ ($7\times10^{-25}$) $\mathrm{cm^3\cdot s^{-1}}$. Interestingly, the behavior around the maximum sensitivity is fairly flat over more than an order of magnitude in DM mass, especially in the case of $b\bar{b}$. We recall that values roughly two orders of magnitude above the thermal relic cross section are ruled out for canonical WIMPs.

We also include in Figure~\ref{fig:limits_bb_tautau} the 68\% containment uncertainty bands on $F_{min}$ for the three observational strategies. In the case of the EXPO method we also include, via quadrature, the uncertainty in both our estimate of the total observed sky area and the average exposure time. Nevertheless, the latter uncertainties turn out to be completely negligible when compared to the $F_{min}$ uncertainty (see \ref{app:CTA_extrapolation}). This reinforces the accuracy of our results: should the actual EXPO time significantly differ from the one we anticipated here, the impact of these variations would still be subdominant in the computation of DM limits when compared to uncertainties in $F_{min}$.

Finally, we remind the reader that, for the computation of these DM limits, we assumed Galactic subhalos to be indeed dark for masses $M_{sub}<10^8 M_{\odot}$. Note that the precise value of this mass cut will directly impact the value of $J_{max}$ and, thus, ultimately, the DM constraints. This particular ansatz is discussed and relaxed in \ref{app:mass_cut}.

\section{Discussion}
\label{sec:discussion}
\begin{figure}[!ht]
\centering
\includegraphics[width=1.0\linewidth]{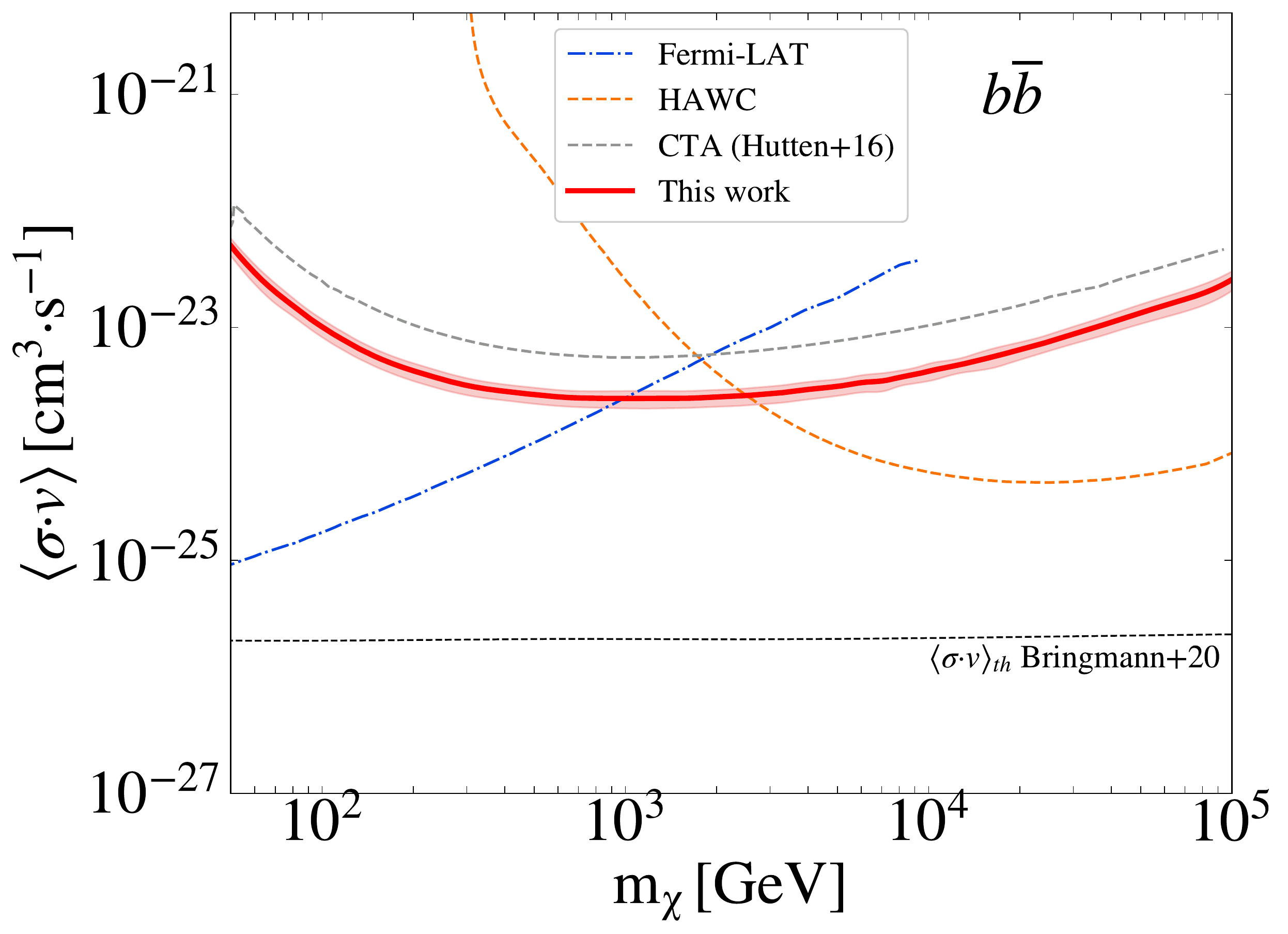}
\vfill
\includegraphics[width=1.0\linewidth]{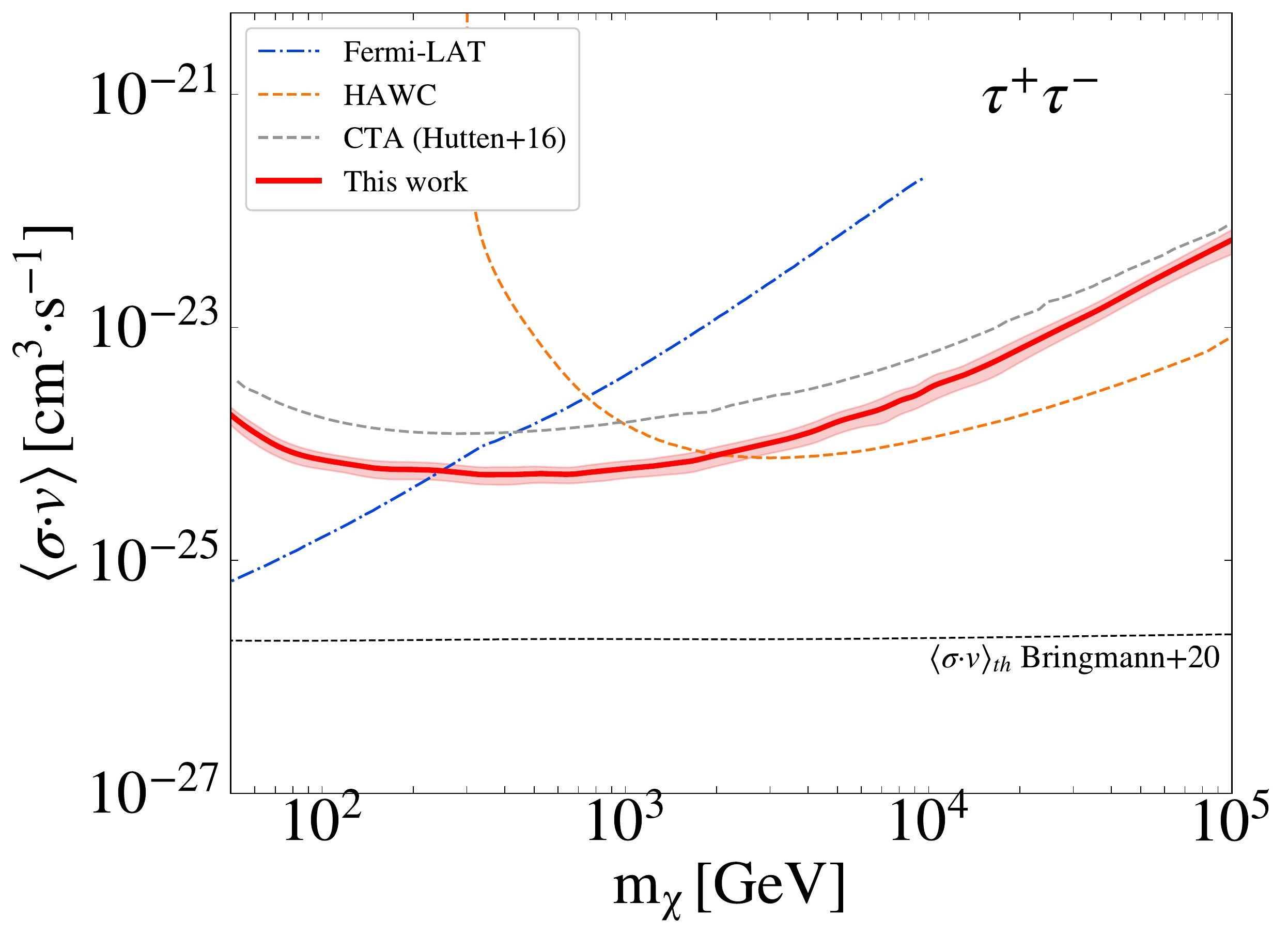}
\caption{Same as Figure \ref{fig:limits_bb_tautau}, but showing only the EXPO constraints and the comparison with previous works, namely the \textit{Fermi}-LAT unIDs \citep{Coronado-Blazquez2019_2}, HAWC unIDs \citep{Coronado-Blazquez2020}, and the previous work on unIDs detection with CTA \citep{Huetten2016}.}
\label{fig:limits_comparison}
\end{figure}

In Figure~\ref{fig:limits_comparison}, we put our results into context by showing a selection of exclusion limits obtained by other instruments that can be compared to ours, as they aim for setting constraints with unidentified sources (unIDs), namely the \textit{Fermi}-LAT \citep{Coronado-Blazquez2019_2}, HAWC \citep{Coronado-Blazquez2020}, and the previous work on unIDs detection with CTA \citep{Huetten2016}. For the sake of clarity, only the best limits, i.e., the ones obtained with the EXPO strategy, are plotted.

The EXPO limits are most competitive for masses above $\sim1$ TeV (500 GeV) for the $b\bar{b}$ ($\tau^+\tau^-$) annihilation channel, of the order of $3\times10^{-24}$ ($7\times10^{-25}$) $\mathrm{cm^3\cdot s^{-1}}$. Our work also improves by a factor $\sim$2 the CTA limits derived in a previous work in this subject by \citet{Huetten2016} -- where authors adopted a different methodology and observation strategy, as well as different CTA IRFs -- being therefore compatible. Note that the behavior around the peak sensitivity is remarkably flat over more than an order of magnitude in the WIMP mass, especially for $b\bar{b}$. We recall that values roughly two orders of magnitude above the thermal relic cross section are ruled out for canonical WIMPs.

In terms of limits, this work also complements previous works on dark subhalo detection for other gamma-ray telescopes, namely the \textit{Fermi}-LAT \citep{Coronado-Blazquez2019_2} and the HAWC \citep{Coronado-Blazquez2020} observatory. Indeed, the synergy between the three mentioned instruments, LAT, HAWC and CTA, is evident: LAT dominates the sensitivity for energies below few hundred of GeV and HAWC above few tens of TeV, with CTA bridging the gap in the region in between. Should a dark subhalo candidate appear in one of the telescopes, this complementarity among instruments would allow for, e.g., follow-up observations with the others. 

The LAT--HAWC--CTA constraints from \cite{Coronado-Blazquez2019_2, Coronado-Blazquez2020} and this work, as shown in Figure \ref{fig:limits_comparison}, have the advantage that they were obtained with the exact same methodology and adopting the same N-body simulation results. This means that these limits are fully consistent one another. Yet, there are important caveats to note: our CTA limits are a projection over a large amount of time. We recall that, in order for the EXPO program to be accomplished, 10 years of CTA operation shall pass. Therefore, the respective weight of the limits at present may change with time: LAT results will probably slowly improve, dominating for long time the constraints below the TeV. In the meantime, HAWC will have several years of data to close the gap from higher energies. CTA will come later and will probably update results over intermediate times (say after ca. 5 years of operation) and possibly extending over 10 years of operation.\footnote{The expected duration of CTA is planned for 30 years.} 

For the moment, we did not include in these discussions at least two players that will enter the scene in the near future: the LHAASO \cite{Sciascio2016} instrument in Tibet and the planned SWGO \cite{Abreu2019} observatory in South America, which might provide stronger constraints than HAWC in the TeV range due to their greater dimensions and thus improved sensitivity.

\bigskip
We do not include in Figure~\ref{fig:limits_comparison} other limits obtained for other DM targets such as the dwarf satellite galaxies of the Milky Way or the Galactic center. There is an ample literature on those results~\citep[see, e.g.,][for a collection of them]{Doro:2014pga}. Both classes of results yield more competitive results than those presented here. Yet, we note that those results are built upon the exact knowledge of the astrophysical factor of the individual target in consideration, and may be therefore subject to larger uncertainties than the results obtained here. For example, the assumption of a cored DM profile in the Galactic center may deteriorate the results by two orders of magnitude, as shown in~\citet{Acharyya:2020sbj}. In this regard, our results have the advantage of being based on conservative assumptions on the statistical properties of the subhalo population as obtained from N-body cosmological simulations, this way being less sensitive to individual target uncertainties.

\section{Summary and conclusions}
\label{sec:conclusions}
In this paper, we assessed the detectability of dark subhalos with the Cherenkov Telescope Array (CTA). We proposed three different observational strategies based on the current Key Science Programs (KSPs) and the predictions for the sky pointings: i) EGAL scenario, based on a planned extragalactic survey that will scan 25\% of the sky (Figure~\ref{fig:egal_map}); ii) DEEP scenario, consisting of a deep-field exposure of 100~h over a region of $10\times10$~deg$^2$; and iii) EXPO scenario, based on the total exposure gathered over 10 years of CTA operation (Figure~\ref{fig:total_exposure_map}).

A careful characterization of the sensitivity of CTA to dark subhalos in these observation scenarios was then performed by means of 100 simulations of a putative subhalo made of WIMPs annihilating via two different annihilation channels ($b\bar{b}$ and $\tau^+\tau^-$). We considered 29 WIMP masses ranging from 50 GeV up to 10 TeV (Figure~\ref{fig:fmin} and Table~\ref{tab:fmin}) and the latest instrument response functions for CTA. 
Based on hundreds of realizations of the Via Lactea II N-body cosmological simulation, ``repopulated'' with subhalos well below the resolution limit of the original simulation, we generated a scaling relation between sky area and maximum (statistically expected) subhalo J-factor achievable in such area (Figure~\ref{fig:jfact_area_scaling}), and computed the expected number of dark subhalos above a certain J-factor in the three different scenarios (Table~\ref{tab:jmin}). In the most optimistic scenario, corresponding to EXPO, we found that on average one dark subhalo ($0.5\pm0.7$) with $J_{min}>10^{20}~\mathrm{GeV^2\cdot cm^{-5}}$ can be expected. Such a value is similar to that of the most competitive dSphs.

However we note that this result is based on a simplified model that does not take into account neither a possible signal spatial extension nor any baryonic-induced effects within the subhalos. The former will probably have negligible effects on our results, while the precise relevance of the second mentioned effect is still subject of debate. Indeed, state-of-the-art hydrodynamical simulations show that baryons in the host halo can largely affect its subhalo population, significantly decreasing the abundance of subhalos especially in the innermost regions of Milky-Way-size halos \cite{Kelley2019, garrison-kimmel, Grand2020}. Yet, these simulations do not resolve subhalos lighter than $\mathrm{\sim 10^6M_{\odot}}$ -- the ones of interest for our work -- thus it remains unclear whether or not in the presence of baryons these tiny subhalos will also be tidally stripped, see e.g. \cite{Bosch2017, Errani2019}.

Assuming instead that no dark subhalo is observed, we were able to compute the 95\% C.L. constraints to WIMP annihilation (Figure~\ref{fig:limits_bb_tautau}). This was done by combining CTA sensitivity results with predictions from our repopulated N-body simulations. The EXPO scenario offers the most competitive results, followed by EGAL and DEEP. In fact, the DEEP strategy turns out to be not optimal for this kind of search, as the required exposure times to be competitive exceed the achievable ones of CTA. Nevertheless, we found interesting to include this scenario as an additional possibility a priori worth exploring. The EXPO limits are of the order of $\langle\sigma v\rangle\sim10^{-24}$ cm$^{-3}$s$^{-1}$ for the $\tau^+\tau^-$ channel and slightly above for the $b\bar{b}$ channel. Thus, they are comparable to those obtained with dwarf satellite galaxies by current IACTs, e.g.~\cite{Abdallah2020}. We placed our limits into a more general context in Figure~\ref{fig:limits_comparison}, by showing results obtained with \textit{Fermi}-LAT and the HAWC observatory for dark subhalos as well. This figure illustrated how nicely the three mentioned experiments are expected to complement one another.

\bigskip
Our study shows that, despite the fact that the chance of serendipitously observing a dark subhalo with CTA is slim (only achievable over a very wide area in the sky and by means of very large exposures), even a no subhalo detection can be used to set competitive DM limits. It is also possible that clear gravitational signatures of dark subhalos in the Galaxy will be found in the near future (see, e.g., the works by \cite{Ibata2020,Buschmann2018,Erkal2016} using gaps in stellar streams), as well as promising subhalo candidates (for instance within the pool of \textit{Fermi}-LAT unidentified sources, as in \cite{Coronado-Blazquez2019_2}). In both cases it may be possible to obtain approximated, if not precise, sky coordinates for potential subhalos. In such scenarios, that should not be discarded, CTA would point to such candidates directly.

We underline that the obtained prospects do not require any specific pointing of the instrument to become a reality. Nonetheless, when CTA starts operating, its performance will most likely not be exactly what is stated in this paper and, thus, these forecasts may require updates. Yet, this paper is meant to already pave the way on the methodology to compute dark-subhalo based DM constraints from CTA observations. It is also possible that intermediate results in this context will be produced by CTA after 5~years and even later on, after 15~years. Over such a long exposure, changes and improvements on the instrumental side and in the description of the dark subhalo scenario may happen. However, we preferred not to speculate further on the impact of these changes. Finally, more realistic and powerful cosmological simulations, including not only DM but also hydrodynamic effects due to baryonic matter at the scale of  dark subhalos, will allow us to be more precise on the characterization of the whole subhalo population in the future, this way making our limits more robust. 

\bigskip

\paragraph{Acknowledgement}
\begin{footnotesize}

This work was conducted in the context of the CTA DMEP Working Group and has gone through internal review by the CTA Consortium.

The authors would like to thank especifically Ana Babic, Dario Hrupec, Sa\u{s}a Mikanovich and Kari Nillson for providing the table of observation time and pointing of MAGIC, Alice Donini, Thomas Gasparetto and Francesco Longo for discussions about the divergent pointing, and the whole MAGIC Collaboration for allowing the use of internal information. The authors would also thank the CTA Consortium, especially Gabrijela Zaharijas and Manuel Meyer for helping through the procedure for non-Consortium papers and Elisa Pueschel, Moritz H\"{u}tten and Yago Ascas\`{i}bar for useful comments about this work.

JCB, MASC and AAS are supported by the {\it Atracci\'on de Talento} contract no. 2016-T1/TIC-1542 granted by the Comunidad de Madrid in Spain, by the Spanish Agencia Estatal de Investigación through the grants PGC2018-095161-B-I00, IFT Centro de Excelencia Severo Ochoa SEV-2016-0597, and Red Consolider MultiDark FPA2017-90566-REDC. We also acknowledge use of the Hydra cluster at the Instituto de F\'isica Te\'orica (IFT), on which numerical computations for this paper took place. JCB also gratefully acknowledges the hospitality of the Dipartimento di Fisica e Astronomia Galileo Galilei of Universit\`a degli Studi di Padova, where part of this work was conducted. The work of AAS was also supported by the Spanish Ministry of Science and Innovation through the grant FPI-UAM 2018.

This research made use of ctools, a community-developed analysis package for Imaging Air Cherenkov Telescope data. ctools is based on GammaLib, a community-developed toolbox for the high-level analysis of astronomical gamma-ray data. Also, this research made use of Python, along with community-developed or maintained software packages, including IPython \cite{Ipython_paper}, Matplotlib \cite{Matplotlib_paper}, NumPy \cite{Numpy_paper} and SciPy \cite{scipy_paper}. This work made use of NASA’s Astrophysics Data System for bibliographic information.

\end{footnotesize}

\paragraph{References}
\bibliographystyle{elsarticle-harv}
\bibliography{biblio.bib}

\clearpage
\appendix

\section{Normal vs. divergent pointing}
\label{app:divergent_pointing}
In the standard CTA observation mode (so-called normal or parallel), all CTA telescopes point to the same sky position. This guarantees maximum performance in terms of sensitivity. However, in this observation mode, the FoV is limited to the maximum angular acceptance of the each telescope camera. 

\begin{figure}[!ht]
\centering
\includegraphics[width=1\linewidth]{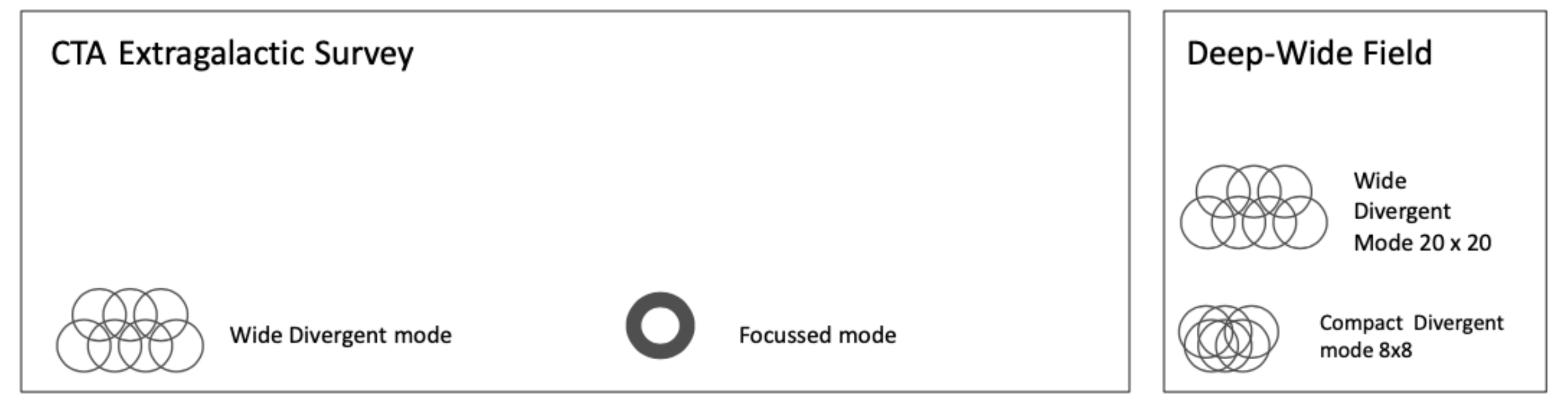}
\caption{Schematic view of different pointing strategies, both for the EGAL and DEEP surveys, in normal and divergent mode.}
\label{fig:pointings}
\end{figure}

In the \textit{divergent} mode there is a small tilt between the lines of sight of the telescopes, that is aimed at enlarging the overall FoV. This comes at the expense of the angular and energy resolutions, as well as of the overall sensitivity. It also adds a level of complexity in the analysis~\citep{Szanecki2015}. The divergent pointing may be especially useful for large surveys, such as the EGAL or the DEEP surveys discussed in this work. Both are schematically shown in Figure~\ref{fig:pointings}, in which possible pointing strategies are highlighted. 

According to \citet{Gerard2015}, the divergent pointing is more optimal than the normal, focused pointing for energies above $\sim$5 TeV. This is because the combination of several telescopes (high multiplicity) ensures a more uniform exposure of the FoV. 
As the energy peaks of the DM spectra under consideration are $\mathrm{E_{peak}^{\tau^+\tau^-}}\sim m_{\chi}/3$ and $\mathrm{E_{peak}^{b\bar{b}}}\sim m_{\chi}/30$ \citep{Cirelli+12}, the divergent would be better than a normal pointing for WIMP masses $\gtrsim$15 TeV for $\tau^+\tau^-$ and $\gtrsim$150 TeV for $b\bar{b}$ (which is already outside the WIMP parameter space due to the unitarity bound) \citep{Smirnov2019, Griest1990}. As we mentioned earlier, for masses larger than $\sim10$ TeV, especially for the case of $\tau^+\tau^-$, the effect of electroweak corrections are unclear and therefore any results above this mass must be taken with caution.

\bigskip
In order to match the normal pointing mode performance by means of adopting the divergent pointing, as in the latter the sensitivity worsens by $\sim$25\% according to \citet{Gerard2015}, the exposure time should increase by $\geq$25\% assuming that the sensitivity scales linearly with time (which may be overly optimistic for low energies, but it is expected to be the case for VHE \citep{Charles2016}). This means almost 4 hours per pointing in the EGAL survey, which is in the limit of feasibility given the typical area of divergent pointing and total time of the survey (1000~h). Therefore, the divergent pointing can only match, at most, the performance of the standard focused one, but not improve it. 

Additionally, \citet{Gerard2015} assumed a number of telescopes that is more similar to the one under consideration for the CTA Southern array (99), while the Northern array is significantly smaller (19) but also participates to the EGAL survey. Indeed, \citet{Gerard2015} shows that the performance decreases significantly for a CTA-North-like configuration. 

As a summary, the divergent pointing may be only as good as the parallel one in the context of this work. Depending on specific assumptions on the array configuration and the scaling of the sensitivity with time, the divergent pointing performance would only worsen with respect to the parallel one.

Yet, a divergent pointing observation comes at hand useful when dealing with extended sources, such as it may happen in the case of dark subhalos, for which the brightest ones are expected to have considerably large angular extensions~\citep{Coronado-Blazquez2019_2}. Although promising thanks to the wide FoV that can be achieved, which may be able to reconstruct the annihilation profile of the source, we note that \citet{Gerard2015} finds a 30\% average loss of angular resolution between 125 GeV and 10 TeV in the divergent pointing compared to the normal pointing.

\section{Building up the CTA EXPO scenario from MAGIC observation records}
\label{app:CTA_extrapolation}
In this Appendix, we explain the details of the extrapolation we did to build the CTA EXPO scenario starting from past MAGIC stereo operations. The latter can be completely characterized by three variables: RA, DEC and exposure time. If several pointings are done to the same (RA, DEC) sky location, we will consider their envelope.

\begin{figure}[!ht]
\centering
\includegraphics[width=0.9\linewidth]{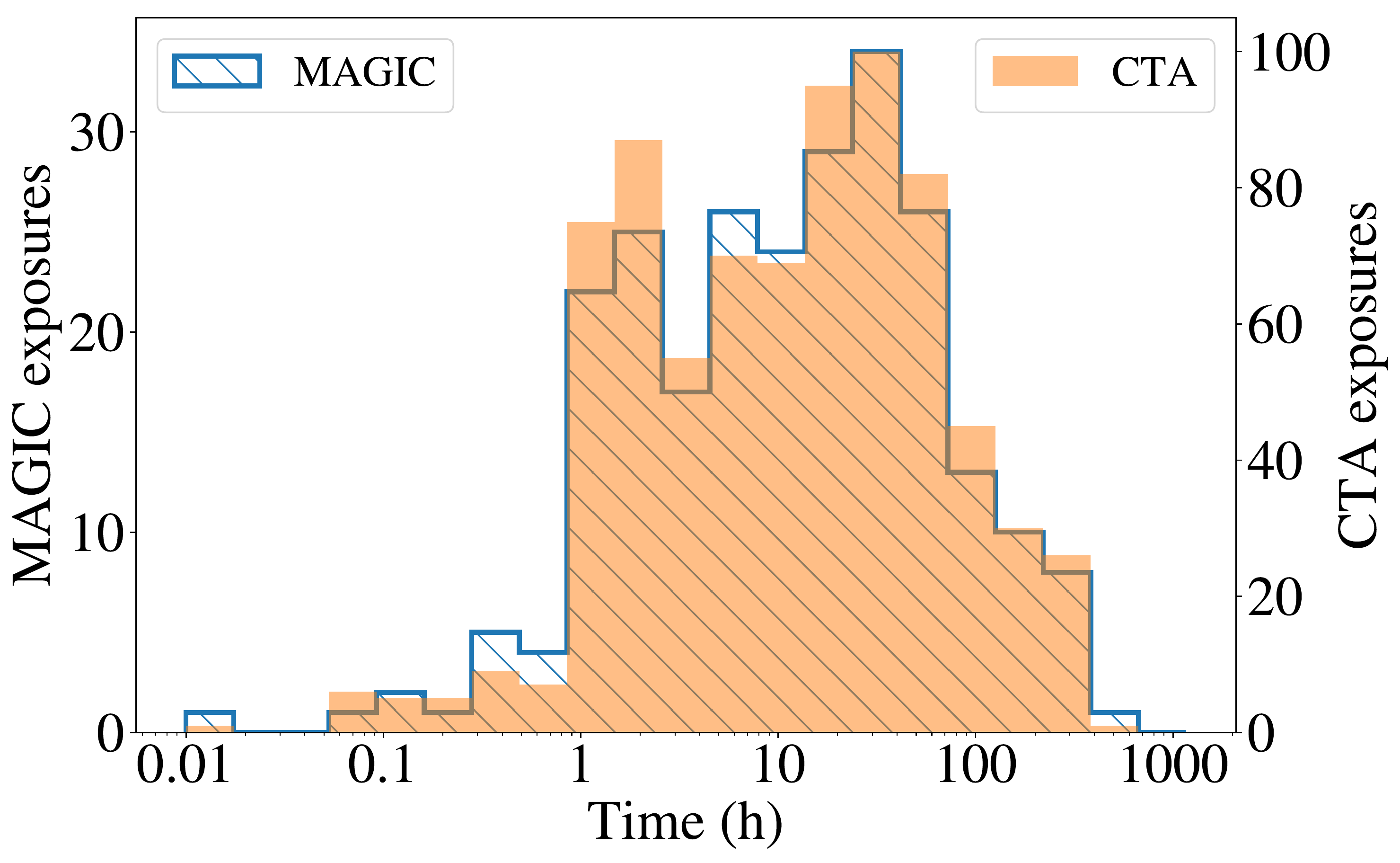}
\vfill
\includegraphics[width=0.9\linewidth]{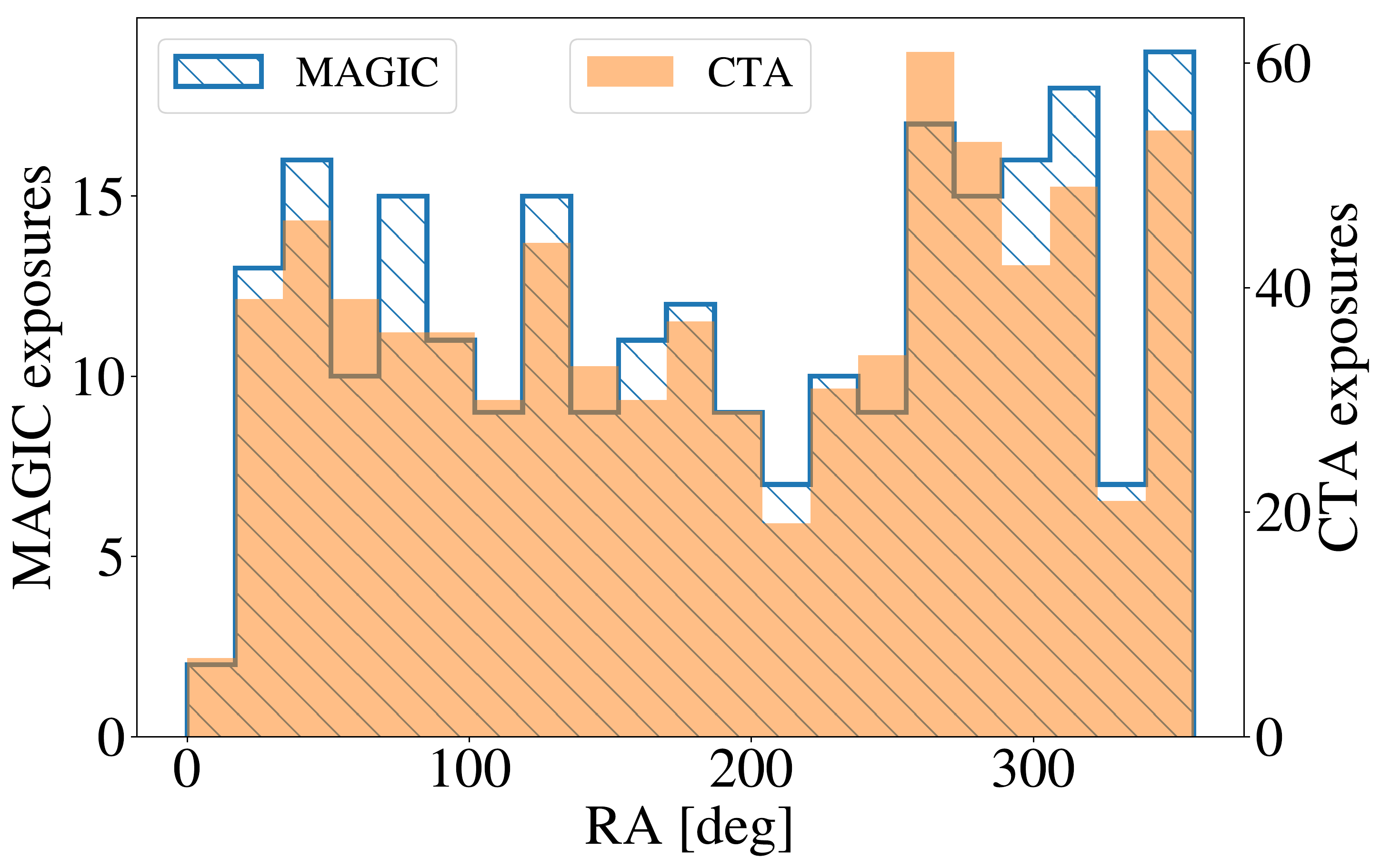}
\vfill
\includegraphics[width=0.9\linewidth]{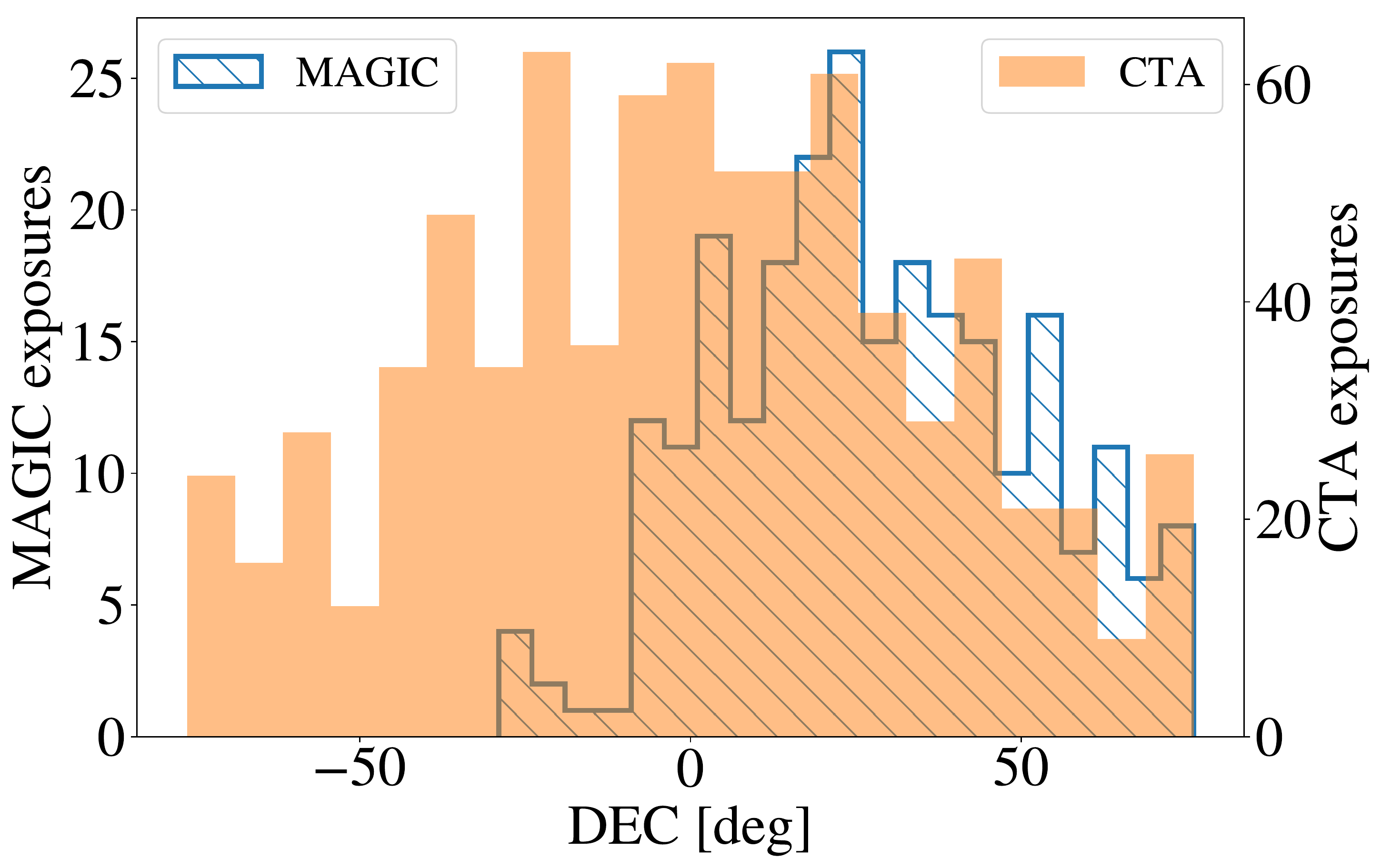}
\caption{For a random realization, this figure shows the distribution of exposures in observation time (upper panel), RA (middle panel) and DEC (bottom panel), both for actual MAGIC stereo operations in 6.5 years and our extrapolation to 10 years of CTA observations. Note that the DEC distributions in the bottom panel must be different, as we assumed half of the observations to be performed by CTA South. See text for details.}
\label{fig:CTA+MAGIC_dist}
\end{figure}

Once the MAGIC distributions of RA, DEC and time are properly characterized, we generate mock random data following the same distributions. For CTA, we extrapolate the number of pointings assuming that they increase linearly with time. From the 250 MAGIC stereo pointings (combining multiple exposures) in 6.5 years, we generate $1.54\times2\times250=770$ observations. The first factor refers to the 10 years of time we assume for CTA, while the factor $2$ comes from the fact of having two arrays, one in each hemisphere. In the case of DEC, we invert the DEC of half of the MAGIC pointings, as we naively assume that half of the observations will be performed by CTA South, which will be located at a latitude $-23.4^\circ$ (quite similar in absolute value to the $28.8^\circ$ of CTA North). Note that, as the distributions are drawn randomly, CTA-South pointings will not be just a mirrored image of those of CTA-North. In Figure \ref{fig:CTA+MAGIC_dist}, we plot both the actual distributions of MAGIC stereo and the resulting CTA extrapolation.

\begin{figure}[!ht]
\centering
\includegraphics[width=1.0\linewidth]{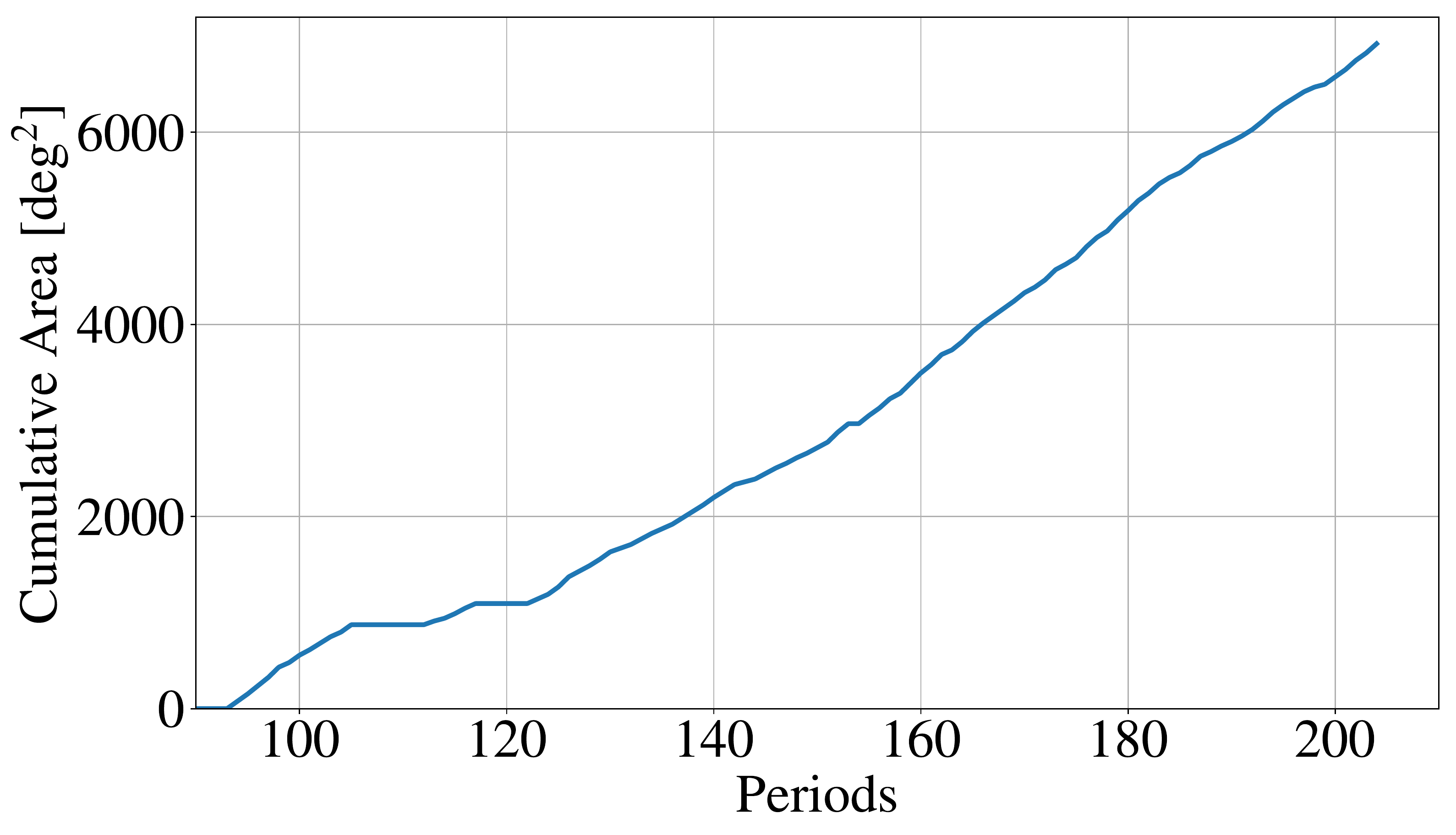}
\caption{Growth of the cumulative exposure area as a function of period of observation (each period is equivalent to 29 days) for MAGIC stereo mode in 6.5 years. Except for some periods where no observations were made, the behavior is linear.}
\label{fig:MAGIC_area_evolution}
\end{figure}

We also checked if the linear extrapolation of pointings with time is a valid approximation. In Figure \ref{fig:MAGIC_area_evolution} we plot the growth of the cumulative exposure area as a function of the periods of observation as done by MAGIC, where each period represents 29 days (moon cycle). The linear regime represents a good approximation, and therefore we can safely extrapolate the number of pointings linearly with time.

To compute the total observed area, we will assume the FoV of all pointings as a circular area of 6 deg of diameter. This number is motivated by the fact the observations will mostly be performed by the MSTs, the most sensitive CTA telescopes in the medium energy range. MSTs have 8 deg of FoV but the off-axis degradation is expected to be non-negligible from 6 deg outwards, thus a value of 6 deg is conservatively adopted.

Also, possible overlaps between observations must be taken into account. The area of intersection between two circles of the same radius $r$ is given by,
\begin{equation}
    A = r^2\cdot\left[q - \mathrm{sin}(q)\right]
\end{equation}
\noindent where $q = 2\cdot \mathrm{acos}\left[d/2r\right]$ and $d$ is the distance between centers. If $d<6$ deg, i.e., two adjacent pointings are at a distance between their centers smaller than their diameter, there will be some overlap. This overlapping area is properly considered (i.e., removed) in the calculation of the total observed area.

\begin{figure}[!ht]
\centering
\includegraphics[width=1.0\linewidth]{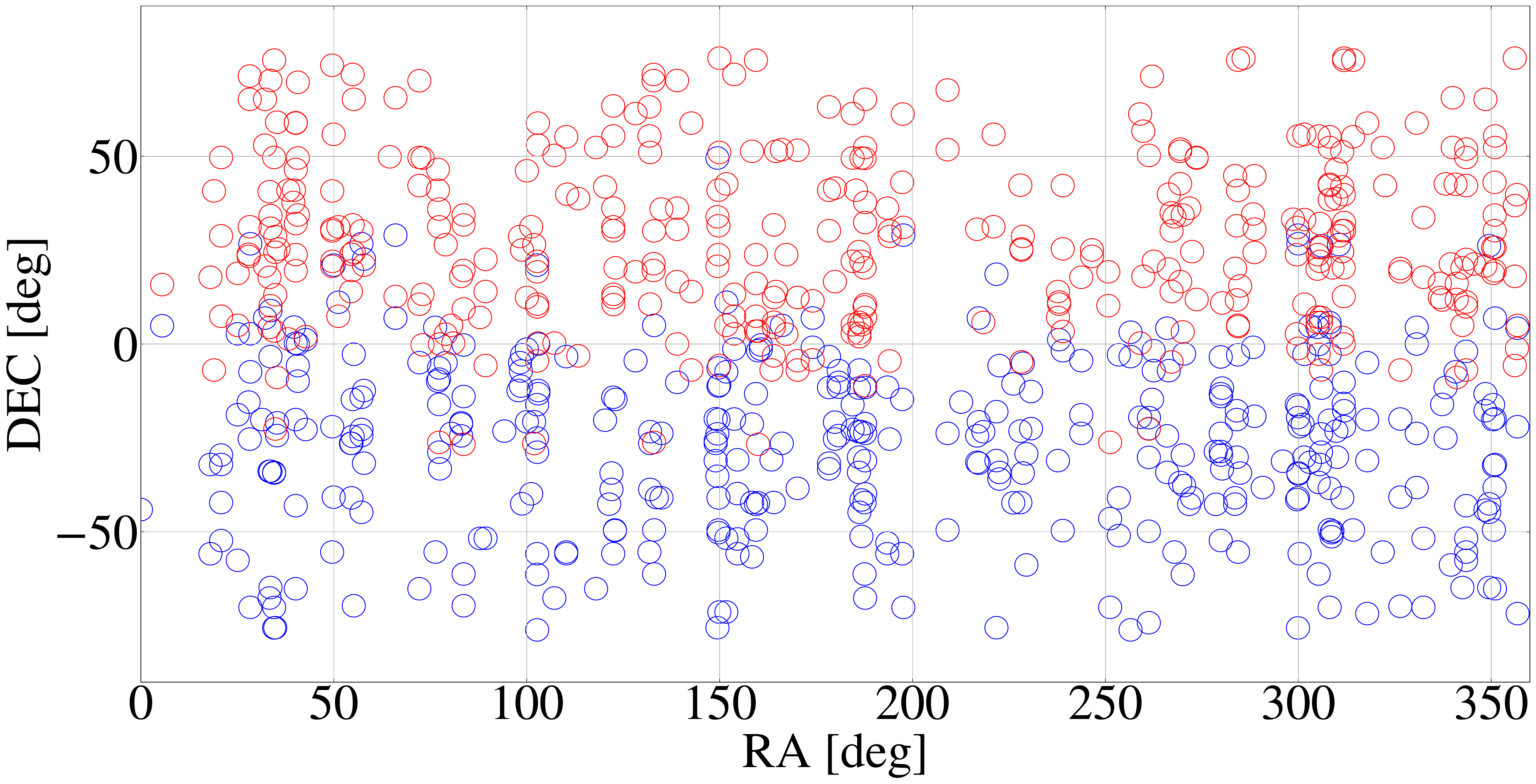}
\caption{Map in equatorial coordinates of CTA pointings in 10 years of operation, extrapolated from 6.5 years of actual MAGIC stereo pointings, as obtained in a single random realization. Red circles are observations performed with CTA North, while blue are for CTA South. The area of each pointing is taken as a circle of 6 deg diameter. Note the overlaps between observations, which have been properly taken into account in the computation of the total effective area of exposure. See text for full details.}
\label{fig:CTA_map}
\end{figure}

Once all this is taken into account, we compute the total area and the average time of observation. To bracket the uncertainties due to the randomness of the map, we generate 2000 realizations. We obtain that the average area surveyed by CTA in 10 years will be $\sim$44\% of the sky, almost doubling that of the EGAL survey, for an average time per pointing of 40h, i.e., a factor $\sim$13 larger than the one that will be adopted for the EGAL survey (3h per pointing)\footnote{The total observation time is around 30000h}. Figure \ref{fig:CTA_map} shows a map of the resulting mock observations. 

\begin{figure}[!ht]
\centering
\includegraphics[width=1.0\linewidth]{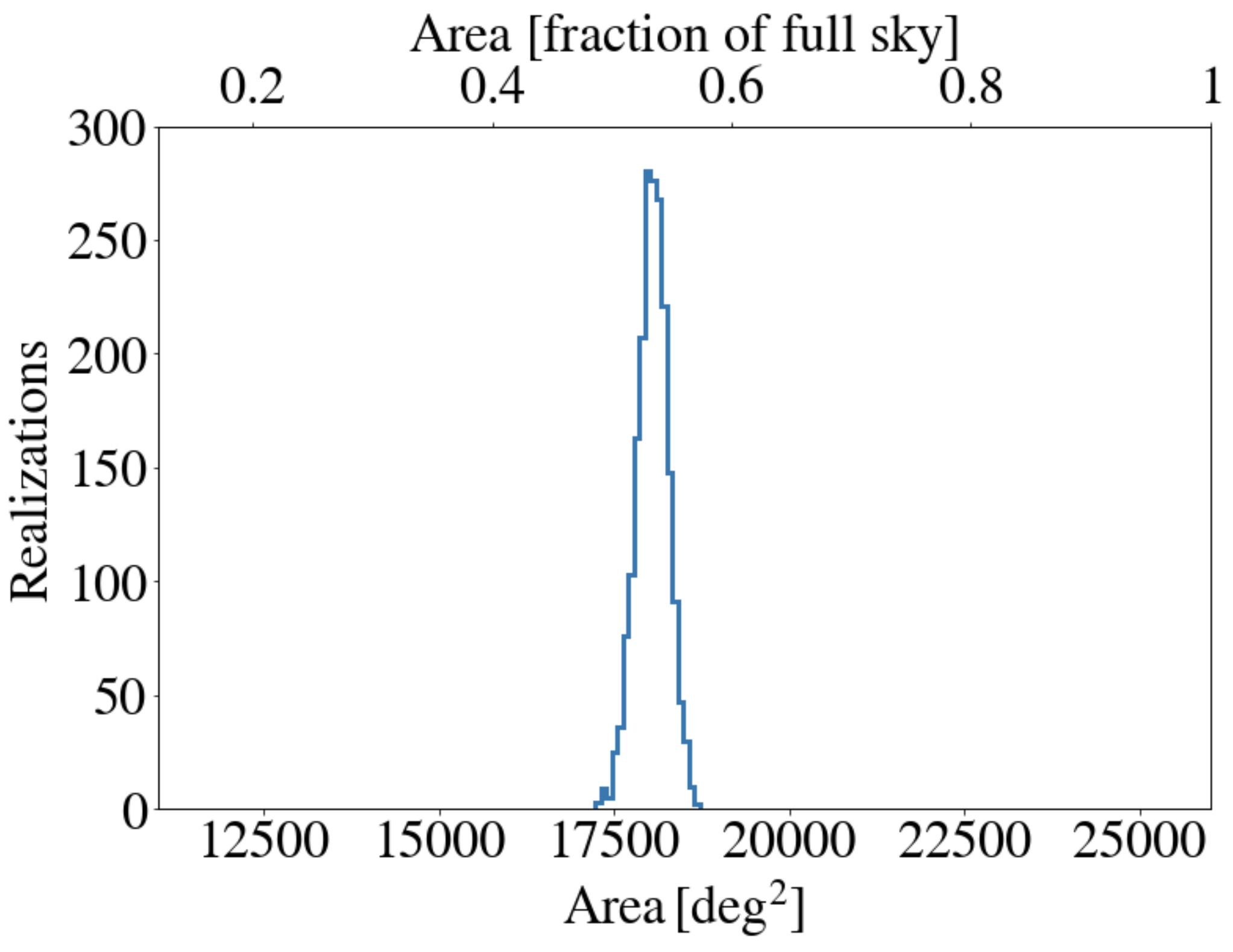}
\vfill
\includegraphics[width=1.0\linewidth]{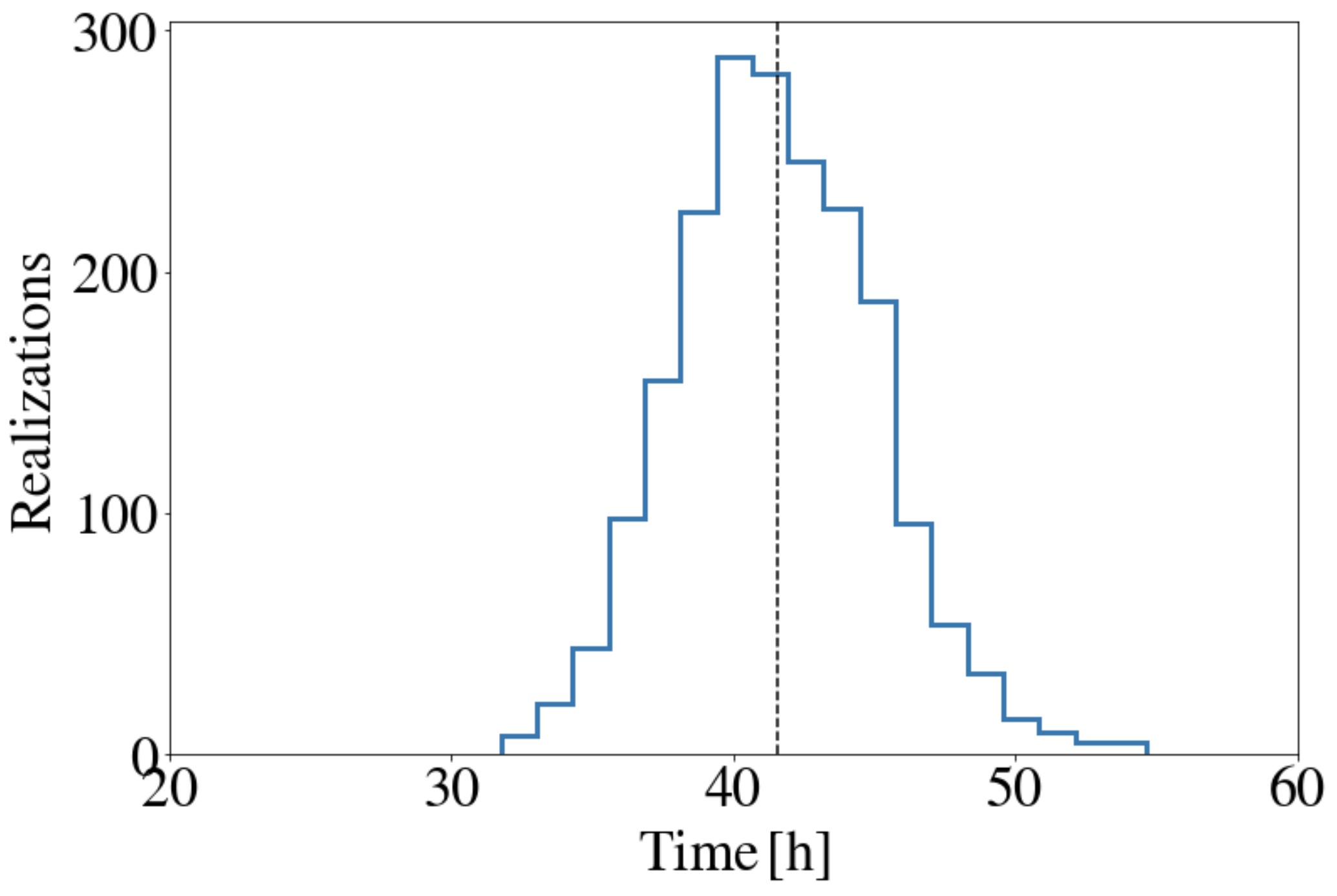}
\caption{Distributions of the total effective area (top panel) and exposure time per pointing (bottom panel) in 10 years of CTA operation, extrapolated from 6.5 years of actual MAGIC stereo observations, as obtained after 2000 random realizations. The vertical line is the average values that we adopt to set constraints in Section \ref{sec:constraints}. The corresponding 1-$\sigma$ uncertainties are below $2\%$ in the case of the area and below $8\%$ in the case of exposure time.}
\label{fig:cta_realiz_dist}
\end{figure}

Finally, we show the full distribution across the 2000 realizations in Figure \ref{fig:cta_realiz_dist}. Both the larger area and mean observation times per pointing will make the EXPO strategy the most competitive one for dark subhalo searches.

\section{Impact of the subhalo mass cut in the dark matter constraints}
\label{app:mass_cut}
\begin{figure}[!ht]
\centering
\includegraphics[width=1.0\linewidth]{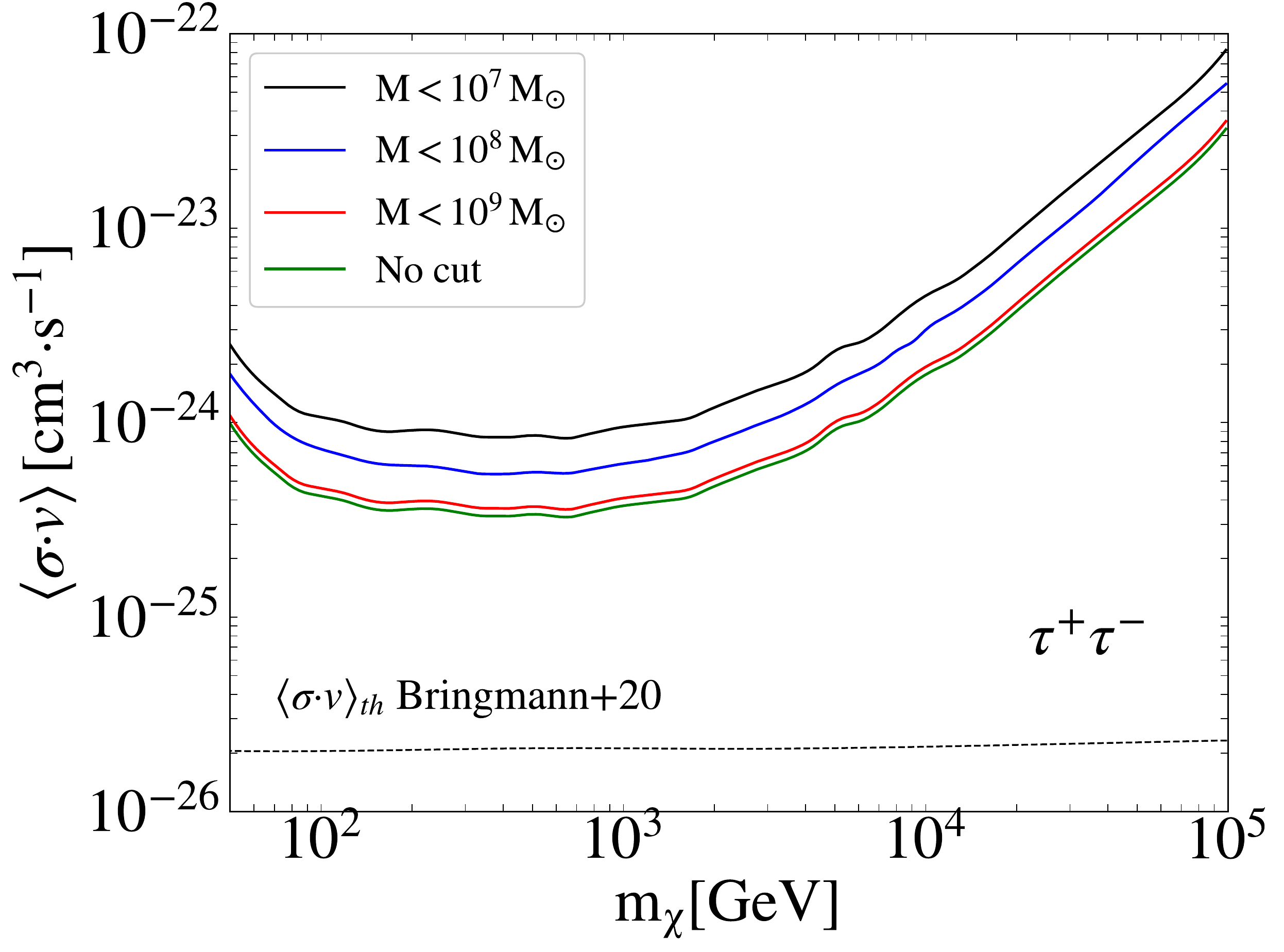}
\caption{EXPO constraints (see Section \ref{sec:constraints} for the $\tau^+\tau^-$ annihilation channel, for different subhalo mass cuts. See discussion in the text for the motivation of the depicted values.}
\label{fig:tautau_masscut}
\end{figure}

To obtain the upper limits on the annihilation cross section shown in Section \ref{sec:constraints}, we assumed a subhalo mass cut of $\mathrm{10^8 M_{\odot}}$, i.e., subhalos with masses larger than this value would host a dSph and would thus be visible, while below the subhalos would be unable to host any significant baryon content and would remain completely dark. Yet, this cut is uncertain as of today~\citep{Gao2004,Zhu2016,garrison-kimmel,Sawala2017,Kelley2019}, and one may ask how dependent the derived constraints on the adopted value are. This is addressed in Figure \ref{fig:tautau_masscut}, where we plot the EXPO constrain for different subhalo mass cuts, as well as no mass cut at all. We note that there are observed dwarf galaxies  with masses as low as few times $\mathrm{10^7 M_{\odot}}$ \cite{Simon2019, Strigari2007, Mateo1998}. Thus, the cases of $\mathrm{M_{cut} = 10^9 M_{\odot}}$ and no-cut shown in this figure are not realistic and are only intended to illustrate the dependence of the constraints with the subhalo cut. On the other hand, a value of $\mathrm{M_{cut} = 10^7 M_{\odot}}$ can be overly conservative, as the precise value at which subhalos go dark is not yet settled, as mentioned above. With all these considerations in mind, we adopted an intermediate yet realistic value of $\mathrm{M_{cut} = 10^8 M_{\odot}}$ as the default one in our limits computation.


From figure \ref{fig:tautau_masscut}, we observe that the DM constraints only vary by a factor $\sim$1.5 for every order of magnitude variation in subhalo mass cut. Also, there is almost no variation for the 'no cut' case compared to the one of $10^{9}$ $M_{\odot}$, as basically the most massive members of the subhalo population are already reached. Therefore, though we adopt a value of $10^8 M_{\odot}$ for the subhalo mass cut as our benchmark model, the constraints can be easily re-scaled for other mass cuts from this figure.


\end{document}